\documentclass[11pt]{article}

\usepackage[final]{acl}

\usepackage{times}
\usepackage{latexsym}

\usepackage[T1]{fontenc}

\usepackage[utf8]{inputenc}

\usepackage{microtype}

\usepackage{inconsolata}

\usepackage{graphicx}

\usepackage{amsmath}
\usepackage{algorithm}
\usepackage{algpseudocode}
\usepackage[most]{tcolorbox}
\usepackage{multirow}
\usepackage{booktabs}
\usepackage{colortbl}
\usepackage{placeins}
\usepackage{caption}
\usepackage{cleveref}
\usepackage{subcaption}
\usepackage{amssymb}
\usepackage{longtable}
\usepackage{array}
\usepackage{makecell}
\usepackage{enumitem}
\usepackage{etoc}

\definecolor{Red}{rgb}{0.768, 0.054, 0.054}
\definecolor{Blue}{rgb}{0.152, 0.294, 0.925}
\definecolor{Green}{rgb}{0,0.4,0.7}
\hypersetup{
    colorlinks=true,
    citecolor=teal,
    linkcolor=Red,
    urlcolor=Green,
}

\newtcblisting{promptbox}[1][]{
  listing only,
  breakable,
  colback=gray!5,
  colframe=black!50,
  boxrule=0.4pt,
  arc=1mm,
  left=1mm,
  right=1mm,
  top=1mm,
  bottom=1mm,
  listing options={
    basicstyle=\ttfamily\footnotesize,
    breaklines=true,
    breakindent=0pt,
    columns=fullflexible,
    keepspaces=true
  },
  #1
}

\definecolor{diffplus}{rgb}{0.00,0.42,0.00}
\definecolor{diffminus}{rgb}{0.70,0.00,0.00}
\definecolor{diffmeta}{rgb}{0.40,0.40,0.40}
\lstdefinelanguage{gitdiff}{
  morecomment=[f][\color{diffmeta}]{@@},
  morecomment=[f][\color{diffplus}]{+},
  morecomment=[f][\color{diffminus}]{-},
  morecomment=[f][\color{diffmeta}\itshape]{//},
}
\newtcblisting{codebox}[1][]{
  listing only, colback=gray!5, colframe=black!50, boxrule=0.4pt, arc=1mm,
  left=1mm, right=1mm, top=1mm, bottom=1mm,
  listing options={language=gitdiff, basicstyle=\ttfamily\scriptsize,
    breaklines=true, breakindent=0pt, columns=fullflexible, keepspaces=true},
  #1
}
\newcommand{\exportentry}[2]{\texttt{#1}~(#2\%)}
\newcolumntype{P}[1]{>{\raggedright\arraybackslash}p{#1}}

\title{Super Library Agent: Joint Generation and Maintenance of Multiple Applications Beyond the Single Codebase}

\author{
  \textbf{Daegyu Sung\textsuperscript{1*}},
  \textbf{Yukyeong Lee\textsuperscript{1*}},
  \textbf{Geon Park\textsuperscript{1}},
  \textbf{Yumin Choi\textsuperscript{1}},
  \textbf{Sung Ju Hwang\textsuperscript{1,2\textdagger}}
  \\
  \\
  \textsuperscript{1}KAIST,
  \textsuperscript{2}DeepAuto.ai
  \\
  \small{
    \texttt{\{sbigstar0310, yukyeong, geon.park, yumin.choi, sungju.hwang\}@kaist.ac.kr}
  }
}

\begin{document}
\maketitle
\etocdepthtag.toc{body}
\begingroup
\renewcommand\thefootnote{}\footnotetext{\textsuperscript{*}\,Equal contribution. \textsuperscript{\textdagger}\,Corresponding author.}
\endgroup
\begin{abstract}
Organizations often develop and maintain \textit{portfolios} of related applications: independently deployable codebases that share substantial domain logic, interface patterns, or operational conventions. 
As LLM coding agents are increasingly used to generate and maintain such software, a naive application-by-application workflow duplicates shared logic across codebases and allows prolonged agentic maintenance to accumulate verbosity, dead code, and structural erosion. 
We introduce the \textbf{Super Library Agent} problem, where an agent sequentially generates a portfolio of $N$ related applications while maintaining a shared \emph{Super Library} of reusable cross-application components. 
A minimal sequential scaffold can in principle extract shared code and migrate applications to the evolving library, but in practice suffers from low extraction recall and fragile dependency migration. 
We address these failures with candidate-guided extraction over code chunk summaries, pre-extraction codebase consolidation, and context-aware migration using extraction traces and call-graph information. 
Across WebGen-Bench and PaperBench, our method preserves application functionality while significantly reducing redundancy and token footprint (verbosity, token length) over zero-shot, and avoiding the structural erosion introduced by naive library construction, with additional reductions in LOC and MDL. Our code is available at \url{https://github.com/sbigstar0310/super-library-agent}.
\end{abstract}

\section{Introduction}

\begin{figure}[t]
  \includegraphics[width=\columnwidth]{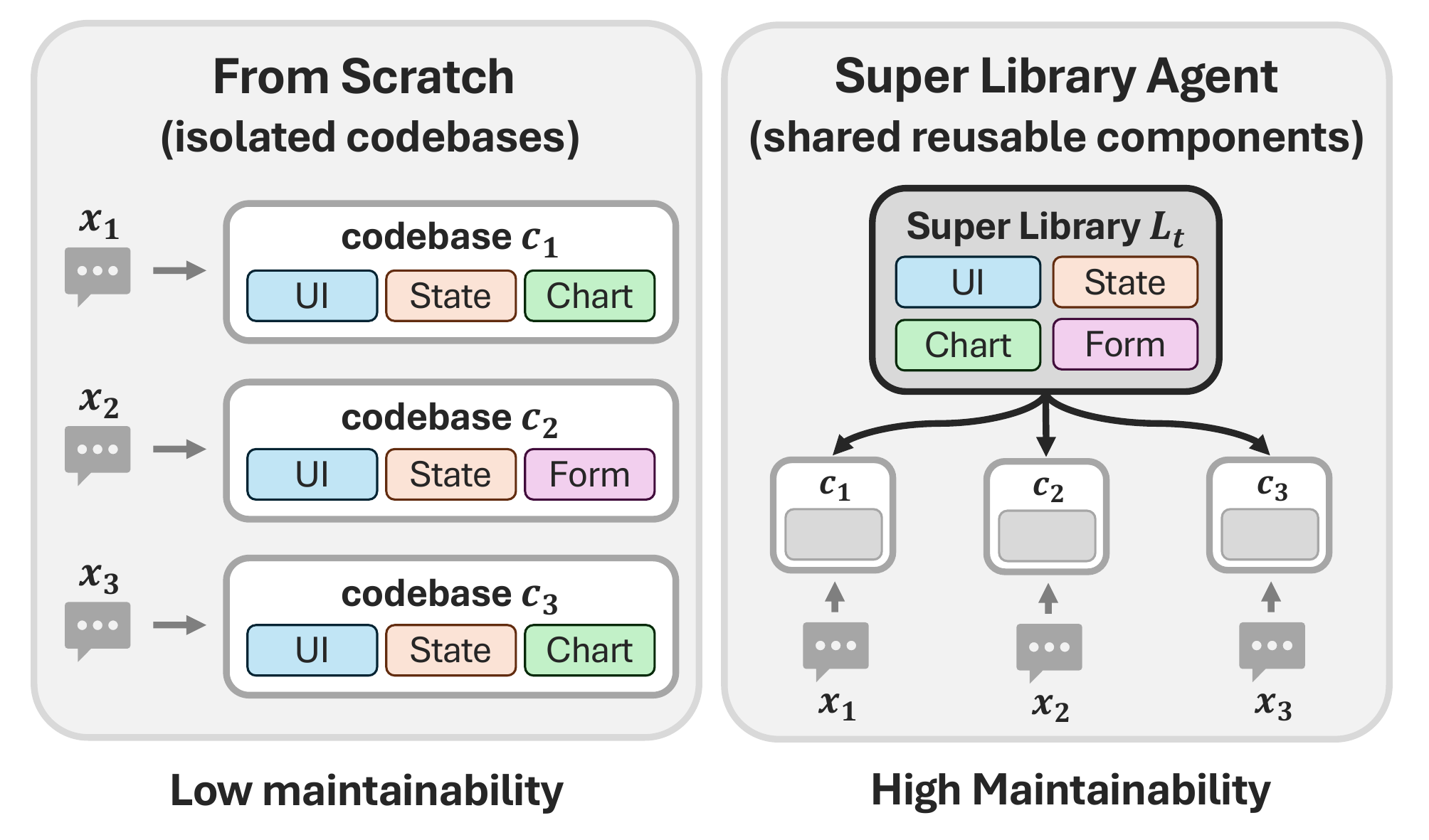}
  \caption{
\textbf{Super Library Agent (SLA) problem.}   Independent generation duplicates shared logic; SLA maintains a shared library across a sequence of codebases.}
  \label{fig:challenge}
  \vspace{-.5cm}
\end{figure}

LLM coding agents can now generate non-trivial applications end-to-end~\cite{qian2024chatdevcommunicativeagentssoftware, hong2024metagptmetaprogrammingmultiagent, dong2024selfcollaborationcodegenerationchatgpt}, but real-world software development rarely produces a single application in isolation. 
Organizations often maintain \textit{portfolios} of related applications~\citep{clements_2001, pohl2005software}: independently deployable codebases that share domain logic, interface patterns, data-processing routines, or operational conventions. 
Generating each portfolio member independently duplicates these shared concerns, violating DRY principles~\citep{thomas_pragmatic_2019} and creating an inconsistency surface where every bug fix, design change, or policy update must be manually propagated across near-duplicates. 
Such duplication is already a well-known maintenance burden in software portfolios~\citep{van_der_zwaag_refactoring_2025, roy2007survey, 5070547}; LLM-generated portfolios make the problem easier to create at scale.

Furthermore, the issue is compounded by the maintenance behavior of LLM coding agents. LLM code generators are known to emit measurable code smells~\citep{siddiq2022empirical, liu2026debtaiboomlargescale}. SlopCodeBench~\citep{orlanski_slopcodebench_2026} shows that prolonged agentic coding can produce \textit{slop}: code that remains functionally correct while becoming increasingly verbose and structurally convoluted. 
Generating and maintaining a portfolio of $N$ applications with $N$ isolated coding agents compounds the effect: each agent re-derives shared logic in subtly different forms, accumulates its own dead code, and drifts further from its siblings with every maintenance cycle. A method that explicitly factors and re-factors shared concerns across the portfolio is therefore an essential structural defense against agent-induced slop accumulation.

\textit{Library learning} with LLMs originated as a way to amortize reasoning and tool-creation costs across related tasks~\citep{ellis2021dreamcoder, bowers2023stitch, berlot-attwell2026library, stengeleskin2024regalrefactoringprogramsdiscover, wang2023voyager, cai2024large}. 
Recent work, such as Librarian~\citep{kovacic_refactoring_2025}, assumes the input files live inside one logical project, and the authors explicitly leave \emph{``large-scale multi-repo library creation''} as an open problem.

We introduce the \textit{Super Library Agent} problem. 
Given a sequence of related application requests, an agent must generate each application while maintaining a shared \textit{Super Library} of components reused across applications. 
A naive baseline generates each application independently, duplicating shared concerns. 
A stronger sequential scaffold generates applications one by one, extracts shared components into the Super Library, and migrates affected applications to use the updated library. 
Although this scaffold captures the intended functionality--maintainability trade-off, it fails in practice for two main reasons.

\paragraph{Challenge 1: Extraction Recall.} 
 The agent must discover reusable components across an expanding portfolio of codebases. This is difficult because functionally equivalent implementations may differ substantially in syntax and structure~\citep{6976121, Saini_2018}. Retrieval mechanisms that rely primarily on surface similarity may therefore overlook valid reuse candidates. As the portfolio grows, these missed candidates increasingly limit the coverage and reuse of the Super Library.

\paragraph{Challenge 2: Migration Correctness.} Extraction is not a local edit. 
Once a local implementation is moved into the Super Library, applications must update imports, call sites, dependent helper functions, and sometimes surrounding APIs. 
A naive migration agent may replace the visible local implementation but miss non-local dependency updates, leaving broken references, duplicated implementations, or dead code behind. Thus, reliable Super Library construction requires both discovering reusable abstractions and safely propagating their use through application codebases.

We propose an augmented Super Library Agent scaffold that targets these two failures. 
For extraction recall, we build summarized code indexes and use an LLM-based selector to propose explicit cross-application reuse candidates, rather than relying only on implicit discovery. 
For migration correctness, we provide the dependency-migration agent with extraction traces and call-graph context, making explicit which local blocks were consolidated, which library components should replace them, and which surrounding call sites or dependencies must be updated.

We evaluate our method on WebGen-Bench~\citep{lu2025webgenbench} and PaperBench~\citep{starace2025paperbenchevaluatingaisability}, reporting accuracy and various code maintainability metrics. Our contributions are:

\begin{itemize}[leftmargin=*, noitemsep, topsep=2pt]
    \item We formulate the Super Library Agent problem, in which an agent generates a sequence of related applications while maintaining a shared library of reusable cross-application components.
    \item  We identify two key obstacles to naive library construction: low extraction recall and fragile dependency migration. We address them with candidate-guided extraction and context-aware migration, respectively.
    \item We evaluate on WebGen-Bench and PaperBench, showing that our method preserves functionality while improving maintainability metrics, library utilization, and abstraction quality over other baselines.
\end{itemize}

\section{Related Work}

\paragraph{Maintainable code generation.}
Recent code generation work increasingly treats maintainability as an explicit objective. MaintainCoder~\citep{wang2025maintaincoder} studies changing requirements, while SlopCodeBench~\citep{orlanski_slopcodebench_2026}, CodeTaste~\citep{codetaste2026}, and Agentic Refactoring~\citep{horikawa2025agenticrefactoring} show that agents can preserve behavior while accumulating verbosity, duplication, and architectural debt, and often perform only local cleanups. These works mainly study a single evolving codebase. In contrast, SLA targets a portfolio of independently deployable applications whose maintainability depends on the joint codebase: the shared library and all applications that import it.

\paragraph{Library learning and post-hoc refactoring.}
Library learning typically mines abstractions from fixed code corpora, from DreamCoder~\citep{ellis2021dreamcoder} and Stitch~\citep{bowers2023stitch} in program synthesis to LILO~\citep{grand2024lilo}'s LLM-assisted interpretable libraries. However,~\citet{berlot-attwell2026library} shows that LLM library-learning gains can be confounded by extra inference budget and that learned artifacts are often rarely reused. \citet{kovacic_refactoring_2025} brings this line closer to software engineering by framing library construction as MDL-minimizing refactoring over existing code snippets. These methods treat library construction as a post-hoc compression or refactoring step over programs that already exist. SLA instead studies an online setting: applications arrive sequentially, the shared library is used when generating each new application, and updated abstractions are migrated back to prior applications.

\paragraph{Common-logic extraction and migration.}
Existing systems extract common logic from completed codebases through corpus-level grouping or explicit refactoring workflows~\citep{7194591, yue2018automaticclonerecommendationrefactoring, pomian2024emassistsafeautomatedextractmethod}. RefAgent~\citep{oueslati2025refagent} decomposes project refactoring into planning, execution, and repair. These systems show that discovery and migration need more structure than direct rewriting, but they do not address online cross-application abstraction: SLA extracts code blocks from natural-language summaries, synthesizes parameterized APIs by turning codebase-specific values into arguments, and records source blocks in an extract map for migration back to affected applications.

\section{Methods}
\label{sec:methods}

\subsection{Problem Definition}
\label{sec:methods:problem-definition}

We introduce the \textbf{Super Library Agent} problem, a novel challenge for coding-specific agentic workflows that jointly generate and maintain \(N\) related applications with a shared \emph{Super Library}. Unlike single-codebase generation, the agent must implement each new application while identifying reusable components across existing codebases and maintaining them through a common library.

Let \(\mathbf{X} = (x_1, \ldots, x_N)\) be a sequence of application implementation requests. The agent starts with an empty Super Library \(\mathcal{L}_0 = \emptyset\). At each step \(t\), given a new request \(x_t\), previous codebases \(\mathbf{C}_{<t} = \{c_1, \ldots, c_{t-1}\}\), and the current library \(\mathcal{L}_{t-1}\), the agent $\mathcal{A}$ produces
\[
(c_t, \mathbf{C}'_{<t}, \mathcal{L}_t) = \mathcal{A}(x_t, \mathbf{C}_{<t}, \mathcal{L}_{t-1}), 
\]
where \(c_t\) is the new codebase, \(\mathcal{L}_t\) is the updated Super Library, and \(\mathbf{C}'_{<t}\) denotes previous codebases patched to use the updated library.

The ideal Super Library contains every component used by at least two implemented applications:
\[  
\mathcal{L}_t^\star =  
\bigg\{  
u \, \bigg\vert  
\sum_{i=1}^{t} \mathbf{1}[\mathrm{use}(u, c_i)] \geq 2  
\bigg\}.  
\]
Thus, application-specific components remain local, while shared components are extracted and consistently reused.

We consider a \emph{sequential portfolio construction setting}, where applications arrive one at a time with a single initial request and no subsequent patches. The goal is to seek \textbf{a favorable Pareto trade-off between the functionality and joint-codebase maintainability} of the final applications \(\mathbf{C}_N\) and Super Library \(\mathcal{L}_N\):
\[  
\max_\mathcal{A}  
\left(  
S_{\mathrm{func}}(\mathbf{C}_N, \mathbf{X}),  
S_{\mathrm{maint}}(\mathbf{C}_N, \mathcal{L}_N)  
\right).  
\]
Here, functionality measures whether applications satisfy their requests, while maintainability measures whether shared components are properly extracted, deduplicated, reused, and updated.

\begin{figure*}[t]
  \includegraphics[width=\linewidth,trim={1cm .5cm 1cm .5cm}]{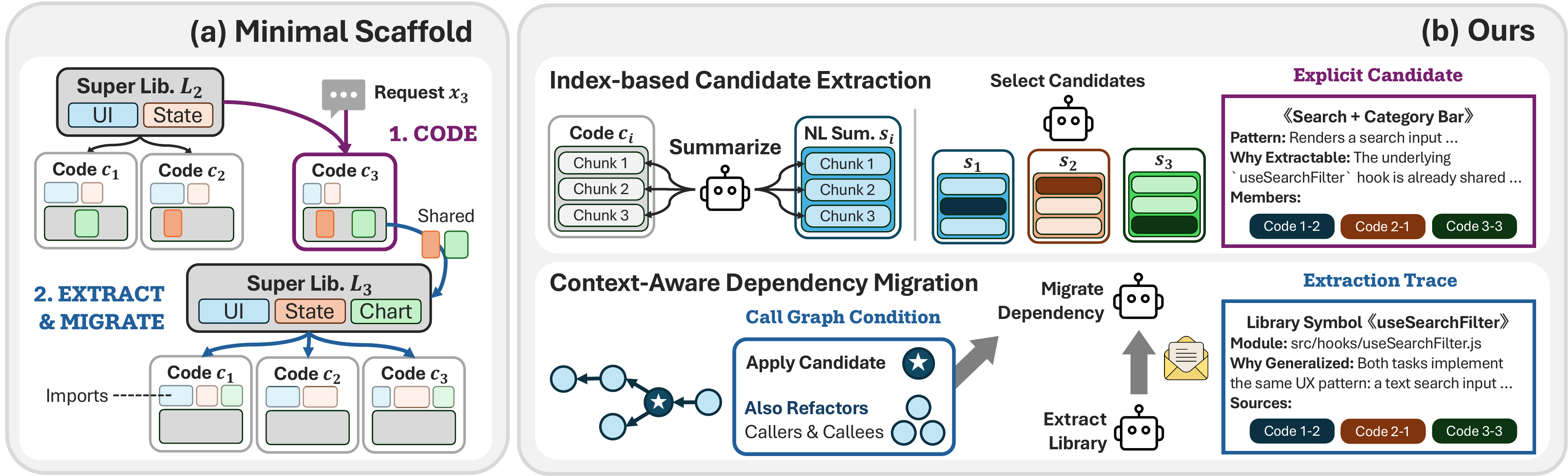}
  \caption{
\textbf{Overview of the proposed workflow.}
Minimal scaffold \textsc{SLA-Naive} (left) vs.\ \textsc{SLA-Full} (right).
  }
  \label{fig:method}
\end{figure*}

\subsection{A Minimal Agentic Scaffold}
\label{sec:methods:minimal-scaffold}

We instantiate Super Library Agent with a minimal sequential scaffold that maps \((x_t, \mathbf{C}_{<t}, \mathcal{L}_{t-1})\) to \((c_t, \mathbf{C}'_{<t}, \mathcal{L}_t)\) at each step \(t\) using two specialized phases: coding and library maintenance.

First, the \emph{coding agent} generates the new codebase \(c_t^0\) from \(x_t\), reusing components from \(\mathcal{L}_{t-1}\) when applicable. Second, a single \emph{library agent} jointly extracts components shared across the implemented applications into the Super Library, yielding \(\mathcal{L}_t\), and migrates previous codebases that contain duplicated local implementations or deprecated library dependencies, yielding \(c_t\) and \(\mathbf{C}'_{<t}\).

This minimal scaffold assigns both Super Library construction and cross-codebase dependency migration to a single general-purpose library agent, without dedicated candidate selection or a structured hand-off between the two tasks. We provide the full procedure in Appendix~\ref{appendix:algorithms}; our experiments in Section~\ref{sec:experiments} use this scaffold as the \textsc{SLA-Naive} comparison baseline.

\subsection{Library Extraction and Dependency Migration Strategies}
\label{sec:methods:full-scaffold}

The minimal scaffold in Section~\ref{sec:methods:minimal-scaffold} has two practical limitations: (i) low recall in identifying reusable components across growing codebases, and (ii) the risk of breaking existing applications when migrating them to newly extracted library components. To address these limitations, we decompose the single library agent into two specialized agents: a \emph{library-extraction agent} responsible for updating the shared library and a \emph{dependency-migration agent} responsible for patching previous codebases. We then equip them with two complementary strategies: (i) index-based candidate extraction and (ii) context-aware dependency migration. See Algorithm~\ref{alg:sla_ours} for our augmented scaffold.

\subsubsection{Index-based Candidate Extraction}
\label{section:index_based_cand_extraction}

Functionally equivalent code blocks often share little surface form, so grouping them by code text or embedding similarity overlooks valid candidates. We therefore summarize each block in natural language and select over these compact \emph{code-summary indexes}, matching blocks by what they do rather than by how they are written, for both library extraction and dependency migration.

\paragraph{Code Block Indexing.}
We maintain an up-to-date code-summary index for every application codebase and the Super Library. Each entry records the location of a semantically meaningful code block and its concise natural-language summary. We identify code blocks using abstract syntax tree (AST) boundaries, such as functions, classes, and modules,\footnote{We use \texttt{cocoindex-code}. \url{https://github.com/cocoindex-io/cocoindex-code}} and generate their summaries with an LLM.\footnote{\texttt{gpt-5.4-nano} with medium reasoning effort is used.}

\paragraph{LLM-Based Candidate Selection.}
A dedicated candidate selector uses these indexes for two matching tasks.\footnote{The candidate selector uses \texttt{deepseek-v4-flash} with high reasoning effort.} The library-extraction agent receives \emph{extraction candidates}, code blocks from multiple applications that implement shared functionality. The dependency-migration agent receives \emph{migration candidates}, library components paired with the application-local blocks they can replace. Detailed prompts and example outputs are provided in Appendix~\ref{appendix:prompt-candidate}.

\paragraph{Pre-Extraction Codebase Consolidation.}
Before cross-application candidate selection, we perform a one-time consolidation of each newly generated codebase. Using the same index-based selection procedure, the agent identifies duplicated or overlapping local implementations and refactors them into shared modules within the application. The consolidated codebase is then indexed for Super Library extraction.

\subsubsection{Context-Aware Dependency Migration}
\label{sec:methods:context-aware}

Dependency migration is challenging because extraction and migration are performed by separate agents, and because replacing a local implementation requires coordinated updates to its import declarations and call sites. To address these challenges, we use extraction traces to bridge the two agents and call-graph context to guide dependent edits.

\paragraph{Extraction-Trace-Based Bridging.}
After each library extraction, the library-extraction agent creates a structured extraction trace for every new or updated Super Library symbol. Each trace records where the symbol came from, why the pattern generalizes across applications, and how to replace the original code with the library API. Passing this trace to the dependency-migration agent bridges the two phases without requiring it to rediscover the correspondence from scratch. Appendix~\ref{appendix:extraction-trace} provides an example.

\paragraph{Call Graph Conditioning.}
For each migration candidate, we attach an \emph{also-refactor} list to the migration prompt containing the relevant import declarations and local caller/callee relationships~\citep{1702621}. This context directs the agent to update dependent call sites and imports while removing obsolete local implementations, reducing broken references, duplication, and dead code~\citep{6227109}.

\section{Experiments}
\label{sec:experiments}

\subsection{Experimental Setup}

\subsubsection{Benchmarks}
\label{sec:experiments:benchmarks}

We evaluate our method on two coding domains: WebGen-Bench and PaperBench. 
To adapt existing single-task coding benchmarks to the Super Library Agent setting, we sample disjoint task sequences of length \(N\) and treat each sequence as an evaluation suite. 
Given a suite \(X=(x_1,\ldots,x_N)\), the agent sequentially implements each task while maintaining a shared Super Library across the generated applications.

\paragraph{WebGen-Bench.}
WebGen-Bench~\cite{lu2025webgenbench} evaluates website generation from natural-language instructions using executable functionality tests and visual-quality judgments. We construct three disjoint 8-task suites. 
Because web applications share UI components, layout primitives, styling patterns, and interaction logic, this domain provides natural reuse opportunities. 
To reduce variance from underspecified visual design, we augment all methods' inputs with the same \emph{layout descriptions} generated from rendered reference pages; all the details are in Appendix~\ref{sec:appendix:webgenbench}.

\paragraph{PaperBench.}
PaperBench~\cite{starace2025paperbenchevaluatingaisability} evaluates whether coding agents can reproduce AI research contributions from ICML 2024 Spotlight and Oral papers.
We use the \emph{Code-Dev} variant and construct five disjoint 4-task suites.
Research implementations share abstractions such as training loops, data pipelines, model wrappers, and evaluation utilities, providing a distinct domain for evaluating library reuse.
Suite details are provided in Appendix~\ref{sec:appendix:paperbench-suites}.

\subsubsection{Compared Methods}
\label{sec:experiments:compared-methods}

All methods use the same backbone LLM, \texttt{deepseek-v4-flash}~\citep{deepseekai2026deepseekv4}, and the \texttt{mini-SWE-agent} harness~\citep{yang2024sweagent}. We distinguish SLA methods, which update the Super Library and migrate dependent applications as requests arrive, from baselines that either omit library construction or perform it only after all applications are complete.

\paragraph{Non-SLA baselines.}
\textsc{Zero-Shot} generates each codebase independently, without a shared library or cross-codebase memory. \textsc{Librarian}~\citep{kovacic_refactoring_2025} applies post-hoc library construction to the completed Zero-Shot codebases. It runs library construction $K{=}8$ times and selects the lowest-MDL candidate among those passing the functionality gate.

\paragraph{SLA methods.}
\textsc{SLA-Naive} uses the minimal scaffold introduced in Section~\ref{sec:methods:minimal-scaffold}. We evaluate two variants using implicit candidate discovery and Ward-clustering-based candidate selection. \textsc{SLA-Full} augments SLA-Naive with index-based candidate extraction, pre-extraction codebase consolidation, extraction traces, and call graph-conditioned migration.

\subsection{Evaluation Metrics}
\label{sec:experiments:metrics}

\subsubsection{Application Functionality}

\paragraph{WebGen-Bench.}
We use the two application-level metrics from WebGen-Bench~\cite{lu2025webgenbench}. 
\textbf{Accuracy (\(\uparrow\))} is the weighted pass rate over executable tests, assigning full credit to fully completed cases and half credit to partially completed ones; test operations are executed by a web-navigation UI agent~\cite{he-etal-2024-webvoyager}.\footnote{The navigation agent is driven by \texttt{gpt-5-mini} with high reasoning effort.}
\textbf{Appearance Score (\(\uparrow\))} is a 0--5 LLM judge score ~\citep{zheng2023judgingllmasajudgemtbenchchatbot} on the rendered entry page, measuring visual quality.\footnote{\texttt{gpt-5-mini} with default reasoning effort is used.}

\paragraph{PaperBench.}
We report the \textbf{Code-Dev Score (\(\uparrow\))}~\citep{starace2025paperbenchevaluatingaisability}, a normalized 0--1 LLM-judge score that checks the generated implementation against paper-specific rubric items.\footnote{\texttt{deepseek-v4-flash} with high reasoning effort is used as main evaluator, and \texttt{gpt-5-mini} with default reasoning effort is used as evaluation report parser.}

\subsubsection{Code Maintainability}

We evaluate maintainability over the final generated codebase, including all applications and the shared library, using five metrics.

\noindent\textbf{Lines of Code (LOC, \(\downarrow\)).}
LOC is the total number of non-empty source lines across all generated applications and the shared library. 

\noindent\textbf{Token Length (Tok, \(\downarrow\)).}
Token Length measures the total number of tokens in all generated applications and the shared library.\footnote{Tokens are counted with the \texttt{Qwen-2.5-7B}~\citep{qwen2025qwen25} tokenizer.}

\noindent\textbf{Minimum Description Length (MDL, \(\downarrow\)).}
MDL~\citep{kovacic_refactoring_2025} measures the negative log probability of the generated code under a reference model \(p_\theta\). 
We use Qwen-2.5-7B~\citep{qwen2025qwen25} and adapt the original file-independent formulation to our setting with a dependency-aware variant:
\begin{equation*}
\begin{aligned}
\mathrm{MDL}(\mathbf{C})
&= -\log p_\theta(\mathcal{L}) \\
&\quad - \sum_{i=1}^{N}\sum_{f \in c_i}
\log p_\theta\!\bigl(f \mid \mathrm{deps}(f)\bigr).
\end{aligned}
\label{eq:metric:mdl}
\end{equation*}
where \(\mathbf{C}\) consists of codebases \(c_1,\ldots,c_N\) and the shared library \(\mathcal{L}\), and \(\mathrm{deps}(f)\) denotes the direct imports of file \(f\). 

\noindent\textbf{Structural Erosion (\(\downarrow\)).}
 Structural Erosion~\citep{orlanski_slopcodebench_2026} measures the share of complexity mass in functions with cyclomatic complexity above 10, weighting each function by its complexity~\citep{mccabe1976complexity} and the square root of its source lines of code.
  
\noindent\textbf{Verbosity (\(\downarrow\)).}
Verbosity~\citep{orlanski_slopcodebench_2026} is the fraction of logical lines of code covered by duplicated code or rule-flagged redundant patterns.

\subsection{Initial Portfolio Construction}
\label{sec:experiments:initial-construction}

\begin{table*}[t]
\centering
\scriptsize
\setlength{\tabcolsep}{4.5pt}
\renewcommand{\arraystretch}{1.08}
\resizebox{0.92\textwidth}{!}{%
\begin{tabular}{lccccccccc}
\toprule
& \multicolumn{2}{c}{\textbf{Functionality}}
& \multicolumn{5}{c}{\textbf{Maintainability}}
& \multicolumn{2}{c}{\textbf{Library Size}} \\
\cmidrule(lr){2-3} \cmidrule(lr){4-8} \cmidrule(lr){9-10}
\textbf{Method}
& Acc. $\uparrow$ & Appr. $\uparrow$
& LOC $\downarrow$ & Tok $\downarrow$ & MDL $\downarrow$ & Eros. $\downarrow$ & Verb. $\downarrow$
& LOC & MDL \\
\midrule
\multicolumn{10}{l}{\textbf{WebGen-Bench}} \\
\addlinespace[1pt]
Zero-Shot
& 76.04 & 3.84
& 9393 & 70364 & 34875 & 0.1032 & 0.1603
& -- & --\\
\textsc{Librarian} ($K{=}1$)
& 75.09 & \textbf{3.90}
& 9202 & 69492 & 34566 & 0.1266 & 0.1558
& 184 & 576\\
\textsc{Librarian} ($K{=}8$)
& 75.78 & 3.89
& 8973 & 68159 & \textbf{33919} & 0.1352 & 0.1472
& 222 & 639\\
\midrule
\textsc{Naive-Implicit}
& 76.95 & 3.89
& 9133 & 69777 & 35103 & 0.1567 & 0.1408
& 447 & 1212\\
\textsc{Naive-Ward}
& 76.07 & \textbf{3.90}
& 8786 & 67774 & 34675 & 0.1250 & 0.1370
& 433 & 1175\\
\rowcolor{gray!12}
\textsc{SLA-Full}
& \textbf{77.21} & 3.86
& \textbf{8552} & \textbf{65633} & 34195 & \textbf{0.0987} & \textbf{0.0994}
& 574 & 1630\\
\addlinespace[1pt]
\textit{$\Delta$ vs.\ Zero-Shot (\%)}
& \textit{$+$1.5} & \textit{$+$0.5}
& \textit{$-$9.0} & \textit{$-$6.7} & \textit{$-$1.9} & \textit{$-$4.4} & \textit{$-$38.0}
& -- & --\\
\midrule
\multicolumn{10}{l}{\textbf{PaperBench}} \\
\addlinespace[1pt]
Zero-Shot
& 0.4687 & --
& 6578 & 68603 & 30342 & 0.2558 & 0.8665
& -- & --\\
\textsc{Librarian} ($K{=}1$)
& 0.4741 & --
& 6608 & 68859 & 30643 & 0.2534 & 0.8628
& 124 & 482\\
\textsc{Librarian} ($K{=}8$)
& 0.4591 & --
& 6551 & 68297 & 30400 & 0.2527 & 0.8628
& 98 & 397\\
\midrule
\textsc{Naive-Implicit}
& 0.4802 & --
& 6445 & 66539 & 29850 & 0.2882 & 0.8582
& 212 & 765\\
\textsc{Naive-Ward}
& 0.4720 & --
& 6440 & 66028 & 29568 & 0.2761 & 0.8645
& 295 & 1027\\
\rowcolor{gray!12}
\textsc{SLA-Full}
& \textbf{0.4809} & --
& \textbf{6252} & \textbf{63514} & \textbf{29489} & \textbf{0.2291} & \textbf{0.8490}
& 336 & 1203\\
\addlinespace[1pt]
\textit{$\Delta$ vs.\ Zero-Shot (\%)}
& \textit{$+$2.6} & --
& \textit{$-$5.0} & \textit{$-$7.4} & \textit{$-$2.8} & \textit{$-$10.4} & \textit{$-$2.0}
& -- & --\\
\bottomrule
\end{tabular}%
}
\caption{\textbf{Initial portfolio construction results.}
Results are averaged over three 8-task WebGen-Bench suites and five 4-task PaperBench suites, each with three trials.
PaperBench Acc.\ denotes the rubric-based Code-Dev Score, and appearance is not applicable.
Maintainability metrics cover the complete portfolio, while Library Size covers only the Super Library.
\textsc{Librarian} $K{=}1$ omits best-of-$K$ selection.}
\label{tab:main-results}
\end{table*}

\paragraph{WebGen-Bench.}
Table~\ref{tab:main-results} (top) shows that all methods achieve comparable functionality, indicating that library use does not harm task performance.
\textsc{SLA-Full} obtains the best scores on LOC, token length, Erosion, and Verbosity, with statistically significant reductions in LOC, token length, and verbosity relative to zero-shot. 
\textsc{Librarian} attains the lowest MDL by selecting the lowest-MDL candidate among $K$ sampled refactorings, but its Erosion rises above zero-shot rather than falling.
In contrast, both \textsc{SLA-Naive} variants reduce some size metrics but increase Erosion, suggesting that unguided library extraction can concentrate complexity into shared components.

\paragraph{PaperBench.}
Table~\ref{tab:main-results} (bottom) shows that \textsc{SLA-Full} achieves the lowest values on all five maintainability metrics while retaining a comparable Code-Dev Score. In contrast, \textsc{Librarian}'s post-hoc MDL selection improves neither MDL nor functionality over Zero-Shot.

\paragraph{Further Evaluation.}
Paired \(t\)-tests show significant improvements in several maintainability metrics, while functionality differences are not statistically significant (Appendix~\ref{sec:appendix:paired}). We further report inference cost and evaluate two cost-reduction strategies (Appendix~\ref{sec:appendix:carryforward}) and compare against recent agentic SE scaffolds (Appendix~\ref{sec:appendix:scaffolds}).

\subsection{Post-Construction Maintenance}
\label{sec:maintenance}

Our primary evaluation measures the functionality and structural maintainability of portfolios after initial construction. Existing benchmarks provide no shared follow-up changes for related applications. We therefore take each final portfolio from Section~\ref{sec:experiments:initial-construction}, consisting of the completed application codebases with or without a Super Library, and apply a shared policy update as an issue patch.

\paragraph{Shared policy updates.}
For each WebGen-Bench suite, we construct one cross-application policy update targeting behavior shared by multiple task specifications. A separate agent authors the update using only the original instructions and layout descriptions, without access to any generated portfolio or library.\footnote{We use \texttt{claude-opus-4.8} with high reasoning effort.} We instantiate the suite-level policy as application-specific issue-patch instructions and corresponding WebVoyager test cases for the target applications. Appendix~\ref{sec:appendix:maintenance} provides the policy construction and evaluation details.

\paragraph{Evaluation protocol.}
We apply the same issue patch to every method using the same edit agent.\footnote{We use \texttt{deepseek-v4-flash} with high reasoning effort.} For methods with a shared library, a single agent jointly updates the library and all target applications. To preserve the independent-codebase setting, Zero-Shot assigns a separate agent to each application and disallows shared-library construction and cross-application imports. We measure \textbf{Patch Size (Added LOC, $\downarrow$)} from the resulting Git diff and report it separately for application and library code. We also evaluate whether the patched applications preserve their original behavior, satisfy the requested behavior, and maintain their visual quality.

\begin{table}[t]
\centering
\small
\setlength{\tabcolsep}{4pt}
\renewcommand{\arraystretch}{1.08}
\resizebox{\columnwidth}{!}{%
\begin{tabular}{lccccccc}
\toprule
& \multicolumn{3}{c}{\textbf{Patch Size}}
& \multicolumn{3}{c}{\textbf{Functionality}} \\
\cmidrule(lr){2-4} \cmidrule(lr){5-7}
\textbf{Method}
& Total $\downarrow$ & App $\downarrow$ & Lib
& Orig. $\uparrow$ & Req. $\uparrow$ & Appr. $\uparrow$ \\
\midrule
Zero-Shot
& 936 & 936 & 0
& \textbf{77.9} & 81.9 & 3.93 \\
\textsc{Librarian}
& 522 & 500 & 22
& 74.0 & 76.9 & 3.83 \\
\midrule
\textsc{Naive-Implicit}
& 632 & 618 & 14
& 76.9 & \textbf{82.9} & 3.82 \\
\textsc{Naive-Ward}
& 380 & 363 & 18
& 77.7 & 81.1 & \textbf{3.95} \\
\rowcolor{gray!12}
\textsc{SLA-Full}
& \textbf{256} & \textbf{232} & 24
& 77.4 & 80.0 & 3.86 \\
\bottomrule
\end{tabular}%
}
\caption{\textbf{Maintenance under shared policy updates.}
Results are averaged over three WebGen-Bench suites and three trials.
Patch Size counts source lines added by the patch.
Orig.\ and Req.\ report post-patch pass rates on the original and requested behaviors.
Appr.\ denotes appearance quality.}
\label{tab:maintenance}
\end{table}

\paragraph{Results.}
Table~\ref{tab:maintenance} shows that \textsc{SLA-Full} achieves the smallest Patch Size overall and across all three suites. By centralizing shared changes in the Super Library, it substantially reduces application-level edits. This realizes the maintainability benefit that motivated our formulation of the Super Library Agent problem. The patched applications retain functionality and visual quality comparable to the baselines.

\begin{table*}[t]
\centering
\begin{subtable}[t]{0.7\linewidth}
\centering
\setlength{\tabcolsep}{3pt}
\renewcommand{\arraystretch}{1.08}
\resizebox{\linewidth}{!}{%
\begin{tabular}{lcccccccccc}
\toprule
& \multicolumn{2}{c}{\textbf{Functionality}}
& \multicolumn{5}{c}{\textbf{Maintainability}}
& \multicolumn{3}{c}{\textbf{Library Size}} \\
\cmidrule(lr){2-3} \cmidrule(lr){4-8} \cmidrule(lr){9-11}
\textbf{Selec.}
& Acc. $\uparrow$ & Appr. $\uparrow$
& LOC $\downarrow$ & Tok $\downarrow$ & MDL $\downarrow$ & Eros. $\downarrow$ & Verb. $\downarrow$
& LOC & Tok & MDL \\
\midrule
None & \textbf{76.95} & 3.89 & 9133 & 69777 & 35103 & 0.1567 & 0.1408 & 447 & 3173 & 1212 \\
Ward & 76.07 & \textbf{3.90} & 8786 & 67774 & 34675 & 0.1250 & \textbf{0.1370} & 433 & 3090 & 1175 \\
NL   & 72.95 & 3.89 & \textbf{8583} & \textbf{66727} & \textbf{32893} & \textbf{0.1122} & 0.1376 & 815 & 5944 & 2057 \\
\bottomrule
\end{tabular}%
}
\caption{Candidate selection for \textsc{SLA-NAIVE}: \textbf{None}, embedding-based \textbf{Ward} clustering, and natural-language code-block summary matching (\textbf{NL}).}
\label{tab:ablation:candidate-selection}
\end{subtable}
\hfill
\begin{subtable}[t]{0.28\linewidth}
\centering
\setlength{\tabcolsep}{2.2pt}
\renewcommand{\arraystretch}{1.08}
\resizebox{\linewidth}{!}{%
\begin{tabular}{l c c >{\columncolor{gray!12}}c}
\toprule
\textbf{Metric}
& \textbf{w/o LC}
& \textbf{w/o CG}
& \textbf{SLA-Full} \\
\midrule
Acc. $\uparrow$    & 75.25  & 74.48  & \textbf{77.21} \\
Verb. $\downarrow$ & 0.1264 & 0.1215 & \textbf{0.0994} \\
Eros. $\downarrow$ & 0.1263 & 0.1409 & \textbf{0.0987} \\
LOC $\downarrow$   & 8733   & 9068   & \textbf{8552} \\
MDL $\downarrow$   & \textbf{33880}  & 34166  & 34195 \\
\bottomrule
\end{tabular}%
}
\caption{\textbf{LC}: Local consolidation, \textbf{CG}: Call graph conditioning.}
\label{tab:ablation}
\end{subtable}
\vspace{-.5em}
\caption{\textbf{Ablations on WebGen-Bench.} We report the mean values over three suites $\times$ three trials. Maintainability metrics are computed over the final portfolio including the shared library.}
\label{tab:ablations}
\end{table*}

\section{Analysis}

Beyond the aggregate results in Section~\ref{sec:experiments}, we analyze how \textsc{SLA-Full} improves reuse and maintainability over sequential application generation.

\subsection{Ablation Study}

In \Cref{tab:ablations}, we compare different candidate selection methods on \textsc{SLA-Naive}, and ablate the major components of \textsc{SLA-Full} to isolate the contribution of each design choice.

\paragraph{Candidate selection for library update.}
In \Cref{tab:ablation:candidate-selection}, we compare three ways of providing reuse candidates to the library-extraction agent in the \textsc{SLA-Naive} setting: no explicit candidates, Ward cluster-based candidates, and our index-based candidate selection.
We observe that our natural language (NL) summary-based approach gives the best scores in four out of five maintainability metrics. This is due to the NL-based approach extracting the largest number of shared components, as seen by the library size. The accuracy degrades in NL due to the additional shared component extraction introducing opportunities for the agent to introduce new bugs. We address this inaccuracy by using call graph conditioning in our \textsc{SLA-Full} approach.

\paragraph{Pre-extraction codebase consolidation.}
We remove the pre-extraction codebase consolidation (\Cref{section:index_based_cand_extraction}) to test whether reducing intra-codebase redundancy before library extraction improves the quality of Super Library candidates. This isolates the effect of filtering duplicated or overly local implementations before they are considered for cross-application extraction.
We observe that without local consolidation, the library agent fails to abstract shared components within a single codebase, since it is designed to extract shared components across multiple codebases. In \Cref{tab:ablation}, this is reflected in the higher Verbosity, Erosion, and LOC metrics compared to \textsc{SLA-Full}. The MDL metric fails to capture this, which is possibly caused by the quality of the training data of the LLM used for evaluating MDL.

\paragraph{Call Graph Conditioning.}
We ablate the migration context by removing call-graph (CG) conditioning, in order to test whether structural caller/callee context is necessary for reliably replacing local implementations with Super Library calls.
In \Cref{tab:ablation}, the accuracy of the no-CG variant is lower, which is due to the dependency-migration agent failing to properly handle the structural dependency updates. Also, without CG, the migration agent often leaves dead code after replacing the local implementation with a Super Library call, which is reflected in higher Verbosity, Erosion, and LOC.

\begin{figure}[t]
  \centering
  \includegraphics[width=1.0\columnwidth,trim={0cm 0.7cm 0 0}]
  {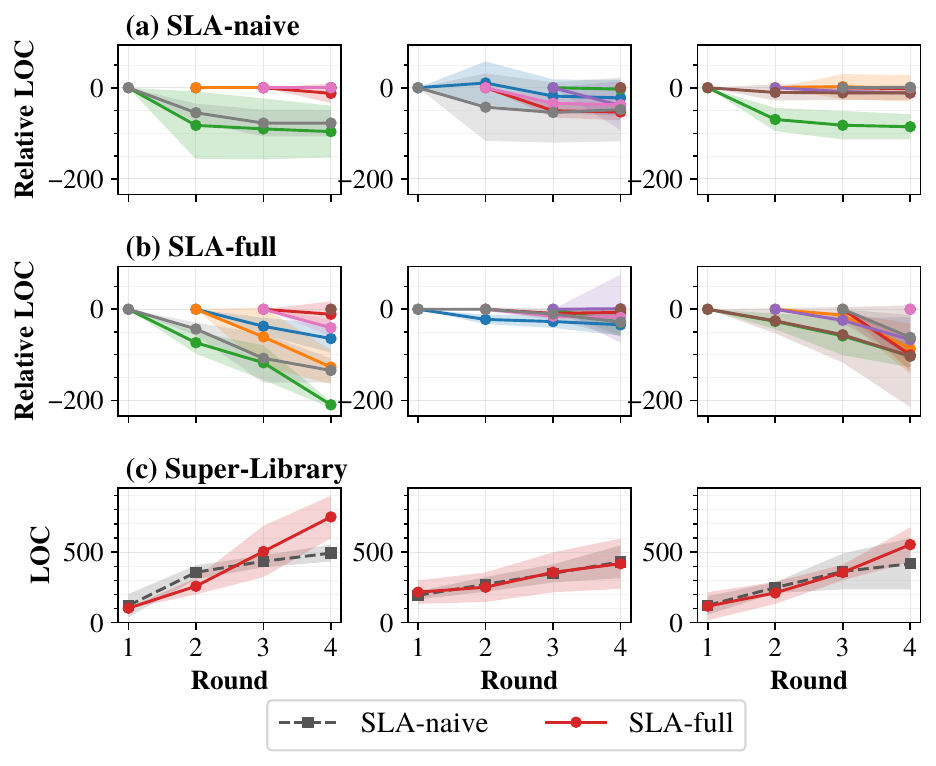}
  \caption{
\textbf{LOC dynamics across sequential rounds. }
Columns are WebGen suites 1--3. In (a) and (b), each colored line is one application. Values are relative to each app's first round; shaded area shows $\pm1$ std over three trials.
  }
  \label{fig:loc-dynamics}
\end{figure}

\begin{figure}[t]
\centering

\begin{subfigure}[t]{\columnwidth}
    \centering
    \includegraphics[width=\columnwidth,trim={0 0 0 1cm}]{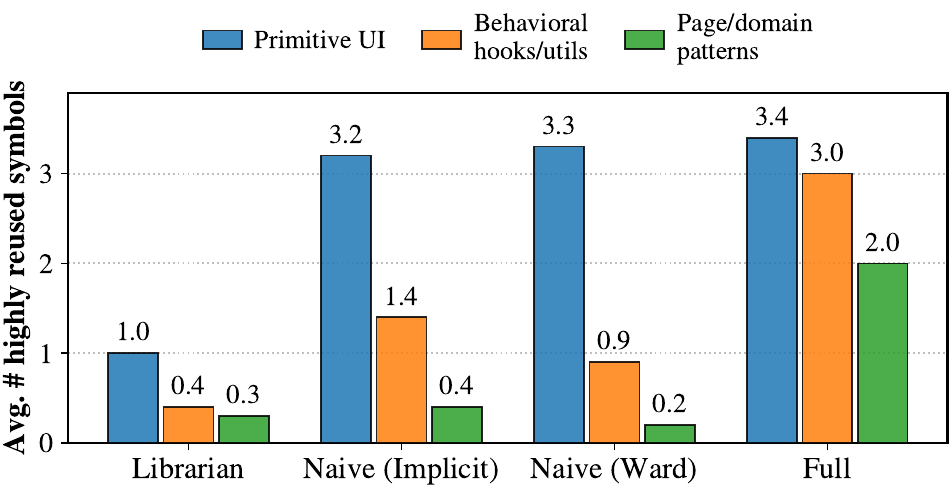}
    \caption{Average number of symbols by category.}
    \label{fig:abstraction-quality-bar}
\end{subfigure}

\vspace{0.75em}

\begin{subfigure}[t]{\columnwidth}
    \centering
    \scriptsize
\setlength{\tabcolsep}{3.5pt}
\renewcommand{\arraystretch}{1.12}
\resizebox{\columnwidth}{!}{%
\begin{tabular}{lccc}
\toprule
\textbf{Method}
& \textbf{Primitive UI}
& \textbf{Behavioral hooks/utils}
& \textbf{Page/domain patterns} \\
\midrule

\textbf{\textsc{Librarian}}
& \begin{tabular}[c]{@{}c@{}}
\texttt{Footer} \\
\texttt{Header}
\end{tabular}
& \begin{tabular}[c]{@{}c@{}}
\texttt{useLocalStorage} \\
\texttt{saveToStorage}
\end{tabular}
& \begin{tabular}[c]{@{}c@{}}
\texttt{ContactForm} \\
\texttt{LoginForm}
\end{tabular}
\\

\midrule
\addlinespace[2pt]

\textbf{\textsc{Naive-Implicit}}
& \begin{tabular}[c]{@{}c@{}}
\texttt{Header} \\
\texttt{Footer}
\end{tabular}
& \begin{tabular}[c]{@{}c@{}}
\texttt{useLocalStorage} \\
\texttt{useFormValidation}
\end{tabular}
& \begin{tabular}[c]{@{}c@{}}
\texttt{ContactForm} \\
\texttt{KpiCard}
\end{tabular}
\\

\midrule
\addlinespace[2pt]

\textbf{\textsc{Naive-Ward}}
& \begin{tabular}[c]{@{}c@{}}
\texttt{Header} \\
\texttt{Card}
\end{tabular}
& \begin{tabular}[c]{@{}c@{}}
\texttt{useLocalStorage} \\
\texttt{createAuthContext}
\end{tabular}
& \begin{tabular}[c]{@{}c@{}}
\texttt{ContactForm}
\end{tabular}
\\

\midrule
\addlinespace[2pt]

\textbf{\textsc{SLA-Full}}
& \begin{tabular}[c]{@{}c@{}}
\texttt{NavigationBar} \\
\texttt{Section}
\end{tabular}
& \begin{tabular}[c]{@{}c@{}}
\texttt{useRouter} \\
\texttt{useFilteredList}
\end{tabular}
& \begin{tabular}[c]{@{}c@{}}
\texttt{FeatureCardGrid} \\
\texttt{MessageBanner}
\end{tabular}
\\

\bottomrule
\end{tabular}%
}
    \caption{Representative symbols (up to two examples per category; full per-run export inventories in Appendix~\ref{appx:abstraction-listing}).}
    \label{tab:abstraction-quality-examples}
\end{subfigure}

\caption{\textbf{Categories of highly reused Super Library symbols} (imported by $\geq4$ of 8 apps). \textsc{SLA-Full} captures more behavioral and page/domain abstractions beyond primitives.}
\label{fig:abstraction-quality}
\end{figure}

\subsection{Maintainability Dynamics Across Rounds}

We track each application's LOC after the last phase of each round. This shows \emph{when} reuse is accumulated, rather than only comparing final scores. In Figure~\ref{fig:loc-dynamics}, \textsc{SLA-Full} tends to reduce app-local code as more applications are added to the stream. In contrast, the naive scaffold shows flatter trajectories, suggesting that a shared library alone is insufficient without reliable candidate discovery and migration.

\subsection{Library Utilization and Abstraction}
\label{sec:analysis:utilization}

We analyze whether the final Super Library is both reused and meaningfully abstracted.

\subsubsection{Library Utilization}

In Table~\ref{tab:library-utilization}, we measure library utilization by counting how many exported Super Library components are reused by different numbers of applications. The Exports column reports all exported library components, while the remaining columns group those exports by the number of generated applications that import them: 1--2, 3--5, and 6--8 apps. The result shows that our \textsc{SLA-Full} builds a larger reusable library rather than merely increasing reuse density over a small set of components. \textsc{SLA-Full} exposes 13.3 components, 5--6 more than the naive variants on average. \textsc{SLA-Full} also increases the number of broadly reused exports, with 6.0 exports used by 3--5 apps and 4.8 exports used by 6--8 apps, compared to 3.3--3.6 and 2.0--2.7 for the naive variants.

\subsubsection{Abstraction Quality}

We analyze what types of abstractions are captured by highly reused Super Library symbols, defined as symbols imported by at least four of eight applications in a run. We categorize them into primitive UI components, behavioral hooks/utilities, and page- or domain-level patterns.

Figure~\ref{fig:abstraction-quality-bar} shows that, apart from \textsc{Librarian}, all methods extract comparable numbers of primitive UI components, but \textsc{SLA-Full} captures substantially more reusable symbols beyond primitives, including behavioral hooks/utilities and page/domain patterns. 
Figure~\ref{tab:abstraction-quality-examples} provides representative examples, showing that the naive variants mostly reuse shallow widgets or thin utilities, whereas \textsc{SLA-Full} additionally extracts interaction/state-management hooks and composed page/domain components.

\begin{table}[t]
\centering
\small
\setlength{\tabcolsep}{4.5pt}
\renewcommand{\arraystretch}{1.08}
\resizebox{0.8\columnwidth}{!}{%
\begin{tabular}{lcccc}
\toprule
 & \multicolumn{3}{c}{\textbf{No. of Apps}}
 & \multirow{2}{*}{\makecell{\textbf{Total}\\\textbf{Exports}}}
\\
\textbf{Method}
& \textbf{1--2}
& \textbf{3--5}
& \textbf{6--8} \\
\midrule

\textbf{\textsc{Librarian}}
& 1.2 & 1.6 & 1.0& 4.2  \\
\textbf{\textsc{Naive-Implicit}}
& 1.3 & 3.6 & 2.7& 8.2  \\
\textbf{\textsc{Naive-Ward}}
& 1.4 & 3.3 & 2.0& 7.2  \\
\textbf{\textsc{SLA-Full}}
& 2.2 & 6.0 & 4.8& 13.3 \\

\bottomrule
\end{tabular}%
}
\caption{\textbf{Super Library utilization,} averaged over three suites $\times$ three trials.}
\label{tab:library-utilization}
\end{table}

\begin{table}[t]
\centering
\small
\setlength{\tabcolsep}{4.5pt}
\resizebox{0.95\linewidth}{!}{%
\begin{tabular}{lcccc}
\toprule
\textbf{Metric} & \textbf{w/o library} & \multicolumn{2}{c}{\textbf{w/ library}} & $\Delta$ \\
\cmidrule(lr){3-4}
                       &        & App & Lib &        \\
\midrule
UI acc. (\%) $\uparrow$        & 80.95    & \multicolumn{2}{c}{84.35}              & $+3.40$    \\
Appear. (0--5) $\uparrow$   & 3.83     & \multicolumn{2}{c}{3.92}               & $+0.09$    \\
\midrule
LOC $\downarrow$               & 8{,}256  & 6{,}567 & 750         & $-940$     \\
MDL $\downarrow$               & 29{,}291 & 27{,}442 & 1{,}985   & $+135$     \\
\bottomrule
\end{tabular}}
\caption{
\textbf{Library-as-prior ablation.} 8 tasks $\times$ 3 trials. $\Delta$ on total (App + Lib).
}
\label{tab:lib_prior}
\end{table}

This suggests summary-guided extraction broadens the library's semantic coverage rather than merely raising primitive counts. Per-suite and per-trial export inventories are in Appendix~\ref{appx:abstraction-listing}.

\subsection{Mature Libraries Enable Richer Application Generation}
\label{sec:lib_prior}

We test whether a Super Library can also serve as a prior for application generation. In this task-informed ablation, a vanilla coding agent generates eight Content-Presentation applications with or without access to \textsc{SLA-Full}'s final library, while extraction, migration, candidate selection, and detailed layout inputs are disabled. Table~\ref{tab:lib_prior} shows that library access improves average accuracy from 80.95 to 84.35 and appearance while reducing application-local code. Even after including the library itself, total LOC decreases by approximately \(11\%\), with only a marginal change in MDL. Figure~\ref{fig:lib_prior_all_c2} shows the per-task comparisons.

\section{Conclusion}

As LLM coding agents build portfolios of related applications at scale, independent generation duplicates shared logic and prolonged agentic maintenance accumulates slop. We formulate this as the Super Library Agent problem, in which an agent develops related applications while maintaining a shared library of reusable components. SLA addresses missed abstractions and incomplete migration through candidate-guided extraction and context-aware dependency migration. Across WebGen-Bench and PaperBench, SLA preserves functionality while improving maintainability, library utilization, and abstraction quality. Post-patch results further show fewer cross-application edits, pointing toward more sustainable LLM coding for larger, evolving codebases.
\clearpage
\section*{Limitations}

\paragraph{Lack of SLA-native benchmarks.}
A major limitation of this work is the absence of benchmarks specifically designed for the Super Library Agent setting. We adapt existing single-task coding benchmarks into sequential suites, but they were not originally built to evaluate long-horizon library growth, cross-application reuse, or codebase maintenance. Section~\ref{sec:maintenance} measures one round of post-deployment maintenance, but two gaps remain: the applications being maintained are themselves benchmark artifacts, generated from task specifications with no users, commit history, or production constraints, so it is untested whether the reduction in patch size survives on real applications; and we measure a single round rather than a sequence, so we do not observe how the shared library holds up as requirements keep drifting or how errors compound across rounds. Future benchmarks should include related application streams with enough shared structure for meaningful reuse, while also testing how the shared library supports evolving maintenance requests.

\paragraph{Limitations of maintainability metrics.}
Our maintainability evaluation relies on proxy metrics such as code size, duplication, verbosity, and library utilization. These metrics capture useful signals about abstraction and reuse, but they do not fully measure whether a codebase is easier to understand, modify, or extend. Moreover, maintainability is ultimately revealed through repeated future changes, whereas our evaluation combines static properties of the final codebases with a single maintenance round. Future work should extend this to multi-round maintenance on real applications, cross-application update consistency, regression frequency, and human judgments of abstraction quality.

\section*{Ethical Considerations}

\paragraph{Automated code modification.}
Our system edits code without a human in the loop: it rewrites existing
applications to import from a shared library, and it applies a maintenance
patch across a whole portfolio in a single pass. A migration that passes our
functional checks can still alter behavior those checks do not cover, and
because one library symbol is imported by many applications, a single faulty
edit propagates to all of them at once. This concentration is the same
property that makes the shared library useful for maintenance, so it cannot
be separated from the method. Applying this to software that people depend on
would require human review of the generated diffs and stronger regression
testing than a benchmark setting provides.

\paragraph{Released artifacts.}
The application codebases and Super Libraries we release are generated from
benchmark task specifications and are intended as evidence for the claims in
this paper, not as deployable software; they have not been reviewed for
security, accessibility, or data-handling practices. The backbone models we
use are trained on public code, so generated components may resemble existing
implementations, and downstream users should verify licensing before reuse.
Our experiments use only public coding benchmarks and involve no human
subjects or personal data (Appendix~\ref{appendix:responsible-nlp}).

\section*{Acknowledgements}

This work was supported by Institute for Information \& communications
Technology Planning \& Evaluation (IITP) grant funded by the Korea government
(MSIT) (RS-2019-II190075, Artificial Intelligence Graduate School Program
(KAIST)), National Research Foundation of Korea (NRF) grant funded by the Korea
government (MSIT) (No. RS-2023-00256259), the Institute of Information \&
Communications Technology Planning \& Evaluation (IITP) with a grant funded by
the Ministry of Science and ICT (MSIT) of the Republic of Korea in connection
with the Global AI Frontier Lab International Collaborative Research
(No. RS-2024-00469482 \& RS-2024-00509279), Institute of Information \&
communications Technology Planning \& Evaluation (IITP) grant funded by the
Korea government (MSIT) (No. RS-2022-II220713, Meta-learning Applicable to
Real-world Problems), and the ``Advanced GPU Utilization Support Program''
funded by the Government of the Republic of Korea (Ministry of Science and ICT).


\bibliography{custom}
\clearpage
\newpage
\appendix
\raggedbottom

\etocdepthtag.toc{appx}
\etocsettagdepth{body}{none}
\etocsettagdepth{appx}{subsection}
\etocstandardlines
\begingroup
\hypersetup{linkcolor=black}
\etocsettocstyle{\section*{Appendix Table of Contents}}{}
\tableofcontents
\endgroup
\clearpage


\section{Initial-Construction Results}
\label{sec:appendix:initial}

\subsection{Paired Statistical Tests}
\label{sec:appendix:paired}

To assess whether the maintainability differences between \textsc{SLA-Full} and each baseline are statistically reliable rather than artifacts of a few suites, we run paired significance tests at the level of individual tasks.
For each metric we form one paired observation per task, taking the unit of pairing to be a (suite, task) pair and averaging over the three trials to reduce run-to-run noise. This yields $n=20$ paired units for PaperBench (five suites of four papers) and $n=24$ for WebGen-Bench (three suites of eight tasks).
For every method pair we compute the mean paired difference $\Delta$ (oriented so that a positive value favors \textsc{SLA-Full} on lower-is-better metrics), a bootstrap 95\% confidence interval over $\Delta$ (10{,}000 resamples), and a two-sided paired $t$-test $p$-value.
\Cref{tab:paired-test} reports the results.
The reductions in token length and Verbosity over zero-shot are significant on both benchmarks, and \textsc{SLA-Full} significantly reduces Erosion relative to the naive baselines. No functionality metric differs significantly between methods, so the maintainability gains do not come at the cost of task performance.
Against \textsc{Librarian}, \textsc{SLA-Full} obtains significant reductions in Erosion and Verbosity on WebGen-Bench and in token length and Verbosity on PaperBench, while no metric significantly favors \textsc{Librarian}.

\begin{table*}[t]
\centering
\scriptsize
\setlength{\tabcolsep}{4pt}
\renewcommand{\arraystretch}{1.15}
\resizebox{\textwidth}{!}{%
\begin{tabular}{l cc cc cc cc}
\toprule
& \multicolumn{2}{c}{\textbf{vs.\ Zero-Shot}}
& \multicolumn{2}{c}{\textbf{vs.\ Librarian}}
& \multicolumn{2}{c}{\textbf{vs.\ Naive-Implicit}}
& \multicolumn{2}{c}{\textbf{vs.\ Naive-Ward}} \\
\cmidrule(lr){2-3}\cmidrule(lr){4-5}\cmidrule(lr){6-7}\cmidrule(lr){8-9}
\textbf{Metric} & $\Delta$\,{\scriptsize[95\% CI]} & $p$ & $\Delta$\,{\scriptsize[95\% CI]} & $p$ & $\Delta$\,{\scriptsize[95\% CI]} & $p$ & $\Delta$\,{\scriptsize[95\% CI]} & $p$ \\
\midrule
\multicolumn{9}{l}{\textbf{WebGen-Bench} ($n=24$)} \\
Acc. $\uparrow$ & -0.85\,{\scriptsize [-3.24, +1.38]} & 0.484 & -1.14\,{\scriptsize [-3.07, +0.78]} & 0.274 & +0.23\,{\scriptsize [-2.38, +2.60]} & 0.864 & -1.04\,{\scriptsize [-3.23, +1.06]} & 0.358 \\
Appear. $\uparrow$ & -0.028\,{\scriptsize [-0.125, +0.042]} & 0.539 & +0.028\,{\scriptsize [-0.069, +0.125]} & 0.575 & +0.028\,{\scriptsize [-0.056, +0.125]} & 0.575 & +0.042\,{\scriptsize [-0.028, +0.125]} & 0.328 \\
LOC $\downarrow$ & +105\,{\scriptsize [+54, +154]} & \textbf{0.001} & +53\,{\scriptsize [+0, +102]} & 0.059 & +73\,{\scriptsize [+28, +126]} & \textbf{0.009} & +29\,{\scriptsize [+3, +56]} & \textbf{0.044} \\
Tok $\downarrow$ & +591\,{\scriptsize [+275, +891]} & \textbf{0.001} & +316\,{\scriptsize [-1, +633]} & 0.068 & +518\,{\scriptsize [+232, +884]} & \textbf{0.006} & +268\,{\scriptsize [+60, +492]} & \textbf{0.028} \\
MDL $\downarrow$ & +85\,{\scriptsize [-43, +206]} & 0.207 & -34\,{\scriptsize [-170, +92]} & 0.623 & +113\,{\scriptsize [-57, +308]} & 0.256 & +60\,{\scriptsize [-77, +207]} & 0.434 \\
Eros. $\downarrow$ & +0.0213\,{\scriptsize [-0.0072, +0.0535]} & 0.189 & +0.1338\,{\scriptsize [+0.0836, +0.1860]} & \textbf{4.5e-05} & +0.1322\,{\scriptsize [+0.0595, +0.2113]} & \textbf{0.002} & +0.1249\,{\scriptsize [+0.0502, +0.2037]} & \textbf{0.004} \\
Verb. $\downarrow$ & +0.0582\,{\scriptsize [+0.0315, +0.0867]} & \textbf{0.001} & +0.0334\,{\scriptsize [+0.0101, +0.0594]} & \textbf{0.017} & +0.0240\,{\scriptsize [-0.0021, +0.0609]} & 0.174 & +0.0455\,{\scriptsize [+0.0251, +0.0719]} & \textbf{0.001} \\
\midrule
\multicolumn{9}{l}{\textbf{PaperBench} ($n=20$)} \\
Score $\uparrow$ & -0.0122\,{\scriptsize [-0.0437, +0.0182]} & 0.458 & -0.0218\,{\scriptsize [-0.0545, +0.0107]} & 0.216 & -0.0007\,{\scriptsize [-0.0242, +0.0274]} & 0.959 & -0.0089\,{\scriptsize [-0.0416, +0.0240]} & 0.607 \\
LOC $\downarrow$ & +81\,{\scriptsize [-15, +165]} & 0.097 & +75\,{\scriptsize [-19, +158]} & 0.120 & +48\,{\scriptsize [-37, +128]} & 0.271 & +47\,{\scriptsize [-56, +146]} & 0.388 \\
Tok $\downarrow$ & +1272\,{\scriptsize [+357, +2071]} & \textbf{0.010} & +1196\,{\scriptsize [+325, +1979]} & \textbf{0.012} & +756\,{\scriptsize [+134, +1369]} & \textbf{0.029} & +628\,{\scriptsize [-543, +1735]} & 0.304 \\
MDL $\downarrow$ & +213\,{\scriptsize [-171, +560]} & 0.273 & +221\,{\scriptsize [-157, +568]} & 0.250 & +90\,{\scriptsize [-203, +345]} & 0.531 & +20\,{\scriptsize [-542, +551]} & 0.945 \\
Eros. $\downarrow$ & +0.0362\,{\scriptsize [-0.0104, +0.0804]} & 0.148 & +0.0291\,{\scriptsize [-0.0157, +0.0705]} & 0.219 & +0.0496\,{\scriptsize [+0.0091, +0.0935]} & \textbf{0.038} & +0.0347\,{\scriptsize [-0.0032, +0.0736]} & 0.101 \\
Verb. $\downarrow$ & +0.0342\,{\scriptsize [+0.0226, +0.0460]} & \textbf{2.5e-05} & +0.0267\,{\scriptsize [+0.0149, +0.0387]} & \textbf{4.0e-04} & +0.0189\,{\scriptsize [+0.0041, +0.0333]} & \textbf{0.022} & +0.0210\,{\scriptsize [+0.0064, +0.0339]} & \textbf{0.008} \\
\bottomrule
\end{tabular}%
}
\caption{\textbf{Paired significance tests} comparing \textsc{SLA-Full} against each baseline (protocol in the text). $\Delta$ is the compared method minus \textsc{SLA-Full}, so \textsc{SLA-Full} is favored by a positive $\Delta$ on lower-is-better ($\downarrow$) metrics and by a negative $\Delta$ on higher-is-better ($\uparrow$) ones; \textbf{bold} marks $p<0.05$. WebGen LOC/Tok/MDL are per-task.}
\label{tab:paired-test}
\end{table*}

\subsection{WebGen-Bench Setup \& Results}
\label{sec:appendix:webgenbench}
\subsubsection{About Layout Descriptions}

For fair comparison, we augment the original instructions with \emph{detailed layout descriptions.}
Because WebGen-Bench instructions are often brief, agents can produce substantially different initial designs, making maintainability metrics sensitive to arbitrary design complexity rather than library reuse. 
We therefore use Claude Opus 4.7 to generate an initial webpage from each raw instruction and describe the rendered layout, then provide this description to all methods together with the original instruction.
This controls design variance and allows us to more directly evaluate code abstraction, migration, and reuse.

\subsubsection{Evaluation Suites}

\Cref{tab:webgenbench-suites} lists the tasks in each suite and how the suites were constructed.

\begin{table}[h]
\centering
\scriptsize
\setlength{\tabcolsep}{4pt}
\renewcommand{\arraystretch}{1.08}
\resizebox{0.95\columnwidth}{!}{%
\begin{tabular}{cl}
\toprule
\textbf{Suite} & \textbf{WebGen-Bench tasks} \\
\midrule
\parbox[t]{6.5em}{\centering S1 \\[1pt]
{\tiny\itshape Content\\ Presentation}\\[1pt]
{\tiny\itshape Brochures \&\\ Portfolios}}
& \begin{tabular}[t]{@{}l@{}}
1. \texttt{[000019]} SW solutions brochure. \\
2. \texttt{[000075]} Team projects \& skills. \\
3. \texttt{[000015]} Clinical office info. \\
4. \texttt{[000016]} Solar company products. \\
5. \texttt{[000040]} Consulting \& training. \\
6. \texttt{[000072]} Personal CV/resume. \\
7. \texttt{[000071]} Designer portfolio. \\
8. \texttt{[000064]} Shipping news blog. \\
\end{tabular} \\
\midrule
\parbox[t]{6.5em}{\centering S2 \\[1pt]
{\tiny\itshape Data\\ Management}\\[1pt]
{\tiny\itshape CRUD / ERP / CRM}}
& \begin{tabular}[t]{@{}l@{}}
1. \texttt{[000085]} Construction project mgmt. \\
2. \texttt{[000022]} Customer-service call log. \\
3. \texttt{[000046]} Hospital admin \& inventory. \\
4. \texttt{[000083]} Detective agency case mgmt. \\
5. \texttt{[000047]} E-government office system. \\
6. \texttt{[000044]} Supply-chain traceability. \\
7. \texttt{[000023]} CRM sales lead tracking. \\
8. \texttt{[000080]} Advanced to-do list app. \\
\end{tabular} \\
\midrule
\parbox[t]{6.5em}{\centering S3 \\[1pt]
{\tiny\itshape User\\ Interaction}\\[1pt]
{\tiny\itshape Social \& Job\\ Platforms}}
& \begin{tabular}[t]{@{}l@{}}
1. \texttt{[000052]} Blue-collar job board. \\
2. \texttt{[000090]} Social networking site. \\
3. \texttt{[000051]} Driver recruitment listings. \\
4. \texttt{[000077]} Engineer tool info site. \\
5. \texttt{[000092]} Skill-sharing platform. \\
6. \texttt{[000053]} Student internship portal. \\
7. \texttt{[000091]} Matrimonial matching site. \\
8. \texttt{[000027]} Promotions \& discounts forum. \\
\end{tabular} \\
\bottomrule
\end{tabular} %
}
\caption{\textbf{WebGen-Bench evaluation suites.} We cluster the 101 WebGen-Bench tasks with Ward agglomerative clustering ($k{=}14$, Euclidean) over \texttt{text-embedding-3-small} embeddings of each task's instruction, category metadata, and UI-test interactions, and select three of the resulting clusters as evaluation suites. From each selected cluster we take the eight tasks closest to the centroid; each suite is then used as a sequential application stream.}
\label{tab:webgenbench-suites}
\end{table}


\subsubsection{Per-Suite Results}

See \Cref{tab:webgenbench-suite-results} for WebGen-Bench per-suite results.


\subsection{PaperBench Setup \& Results}

\subsubsection{Agent Environment}
\label{sec:appendix:paperbench-env}

We pre-install 27 packages in the agent container: PyTorch, NumPy, SciPy,
scikit-learn, and pandas, along with common plotting, I/O, and configuration
utilities. Domain frameworks such as \texttt{stable-baselines3},
\texttt{transformers}, and \texttt{timm} are left out on purpose. Such a
framework already implements a paper's method, so it would hand the agent the
very code that a shared library is meant to capture. We would then be measuring
whether a method calls the framework, not how it organizes its own code. The
container also runs with \texttt{--network=none}, so no agent can install
anything mid-run. The list of available packages is given to the agent in its
prompt. Every method runs in this same container, Zero-Shot included.

\subsubsection{Evaluation Suites}
\label{sec:appendix:paperbench-suites}

\Cref{tab:paperbench-suites} lists the papers in each suite and how the suites were constructed.

\begin{table}[h]
\centering
\scriptsize
\setlength{\tabcolsep}{4pt}
\renewcommand{\arraystretch}{1.08}
\resizebox{0.95\columnwidth}{!}{%
\begin{tabular}{cl}
\toprule
\textbf{Suite} & \textbf{PaperBench tasks} \\
\midrule
S1
& \begin{tabular}[t]{@{}l@{}}
1. \texttt{fre}: Functional Reward Encodings. \\
2. \texttt{rice}: RICE. \\
3. \texttt{sapg}: Split and Aggregate Policy Gradients. \\
4. \texttt{ftrl}: RL Fine-tuning as Forgetting Mitigation. \\
\end{tabular} \\
\midrule
S2
& \begin{tabular}[t]{@{}l@{}}
1. \texttt{robust-clip}: Robust CLIP. \\
2. \texttt{sample-specific-masks}: Sample-specific Masks. \\
3. \texttt{lca-on-the-line}: OOD Generalization. \\
4. \texttt{test-time-model-adaptation}: Test-Time Adaptation. \\
\end{tabular} \\
\midrule
S3
& \begin{tabular}[t]{@{}l@{}}
1. \texttt{mechanistic-understanding}: DPO and Toxicity. \\
2. \texttt{what-will-my-model-forget}: Forgotten Examples. \\
3. \texttt{stay-on-topic-with-classifier-free-guidance}: CFG. \\
4. \texttt{adaptive-pruning}: Adaptive Pruning and Tuning. \\
\end{tabular} \\
\midrule
S4
& \begin{tabular}[t]{@{}l@{}}
1. \texttt{bridging-data-gaps}: Diffusion Transfer Learning. \\
2. \texttt{sequential-neural-score-estimation}: Neural Score Estimation. \\
3. \texttt{stochastic-interpolants}: Data-Dependent Couplings. \\
4. \texttt{all-in-one}: Simulation-Based Inference. \\
\end{tabular} \\
\midrule
S5
& \begin{tabular}[t]{@{}l@{}}
1. \texttt{bam}: Batch-and-Match Variational Inference. \\
2. \texttt{pinn}: Training PINNs Loss Landscapes. \\
3. \texttt{lbcs}: Refined Coreset Selection. \\
4. \texttt{bbox}: BBox-Adapter for Black-Box LLMs. \\
\end{tabular} \\
\bottomrule
\end{tabular} %
}
\caption{\textbf{PaperBench evaluation suites.} We construct five disjoint suites of four tasks each. Suites S1--S3 are topically coherent (RL, VLM, NLP), S4 is a diffusion/score-based inference suite, and S5 is a heterogeneous low-reuse suite included as a robustness probe. Each suite is used as a sequential task stream for evaluating Super Library Agent methods.}
\label{tab:paperbench-suites}
\end{table}


\subsubsection{Per-Suite Results}

See \Cref{tab:paperbench-suite-results} for PaperBench per-suite results.

\begin{table*}[p]
\centering
\scriptsize
\setlength{\tabcolsep}{4pt}
\renewcommand{\arraystretch}{1.05}
\resizebox{0.8\textwidth}{!}{%
\begin{tabular}{llccccc}
\toprule
\textbf{Metric}
& \textbf{Suite}
& \textbf{Zero}
& \textbf{Librarian}
& \multicolumn{2}{c}{\textbf{Naive}}
& \textbf{Full} \\
\cmidrule(lr){5-6}
&
&
&
& {\tiny \textbf{Implicit}}
& {\tiny \textbf{Ward}}
& \\
\midrule

\multicolumn{7}{l}{\textbf{Functionality}} \\

\multirow{4}{*}{Acc. $\uparrow$}
& S1   & 82.99\,{\scriptsize$\pm$2.12} & 81.97\,{\scriptsize$\pm$1.56} & 83.67\,{\scriptsize$\pm$0.00} & 83.67\,{\scriptsize$\pm$3.68} & 83.33\,{\scriptsize$\pm$2.12} \\
& S2   & 70.28\,{\scriptsize$\pm$2.55} & 70.83\,{\scriptsize$\pm$2.21} & 70.83\,{\scriptsize$\pm$3.00} & 70.00\,{\scriptsize$\pm$4.64} & 72.22\,{\scriptsize$\pm$3.37} \\
& S3   & 74.85\,{\scriptsize$\pm$0.52} & 74.55\,{\scriptsize$\pm$1.82} & 76.36\,{\scriptsize$\pm$1.57} & 74.54\,{\scriptsize$\pm$3.96} & 76.06\,{\scriptsize$\pm$2.62} \\
\rowcolor{gray!12}
& Avg. & 76.04\,{\scriptsize$\pm$5.82} & 75.78\,{\scriptsize$\pm$5.17} & 76.96\,{\scriptsize$\pm$5.83} & 76.07\,{\scriptsize$\pm$7.00} & \textbf{77.21}\,{\scriptsize$\pm$\textbf{5.44}} \\

\addlinespace[1pt]
\multirow{4}{*}{Appear. $\uparrow$}
& S1   & 3.84\,{\scriptsize$\pm$0.08} & 3.96\,{\scriptsize$\pm$0.07} & 4.00\,{\scriptsize$\pm$0.00} & 3.96\,{\scriptsize$\pm$0.07} & 3.92\,{\scriptsize$\pm$0.07} \\
& S2   & 3.92\,{\scriptsize$\pm$0.07} & 3.84\,{\scriptsize$\pm$0.08} & 3.84\,{\scriptsize$\pm$0.08} & 3.96\,{\scriptsize$\pm$0.07} & 3.96\,{\scriptsize$\pm$0.07} \\
& S3   & 3.75\,{\scriptsize$\pm$0.00} & 3.88\,{\scriptsize$\pm$0.13} & 3.84\,{\scriptsize$\pm$0.08} & 3.79\,{\scriptsize$\pm$0.08} & 3.71\,{\scriptsize$\pm$0.08} \\
\rowcolor{gray!12}
& Avg. & 3.84\,{\scriptsize$\pm$0.09} & 3.89\,{\scriptsize$\pm$0.10} & 3.89\,{\scriptsize$\pm$0.10} & \textbf{3.90}\,{\scriptsize$\pm$\textbf{0.10}} & 3.86\,{\scriptsize$\pm$0.13} \\

\midrule
\multicolumn{7}{l}{\textbf{Maintainability}} \\

\multirow{4}{*}{LOC $\downarrow$}
& S1   & 7129\,{\scriptsize$\pm$235}  & 6623\,{\scriptsize$\pm$302}  & 6809\,{\scriptsize$\pm$83}   & 6459\,{\scriptsize$\pm$42}   & 6299\,{\scriptsize$\pm$245} \\
& S2   & 9834\,{\scriptsize$\pm$463}  & 9437\,{\scriptsize$\pm$705}  & 9890\,{\scriptsize$\pm$741}  & 9668\,{\scriptsize$\pm$768}  & 9689\,{\scriptsize$\pm$576} \\
& S3   & 11215\,{\scriptsize$\pm$51}  & 10858\,{\scriptsize$\pm$207} & 10700\,{\scriptsize$\pm$734} & 10232\,{\scriptsize$\pm$665} & 9669\,{\scriptsize$\pm$445} \\
\rowcolor{gray!12}
& Avg. & 9392\,{\scriptsize$\pm$1819} & 8973\,{\scriptsize$\pm$1909} & 9133\,{\scriptsize$\pm$1853} & 8787\,{\scriptsize$\pm$1834} & \textbf{8552}\,{\scriptsize$\pm$\textbf{1733}} \\

\addlinespace[1pt]
\multirow{4}{*}{MDL $\downarrow$}
& S1   & 31474\,{\scriptsize$\pm$1691} & 30378\,{\scriptsize$\pm$1886} & 30556\,{\scriptsize$\pm$136}  & 29927\,{\scriptsize$\pm$614}  & 30072\,{\scriptsize$\pm$1193} \\
& S2   & 36138\,{\scriptsize$\pm$3020} & 35371\,{\scriptsize$\pm$3372} & 37607\,{\scriptsize$\pm$2778} & 38014\,{\scriptsize$\pm$2299} & 36924\,{\scriptsize$\pm$3397} \\
& S3   & 37013\,{\scriptsize$\pm$553}  & 36008\,{\scriptsize$\pm$1281} & 37144\,{\scriptsize$\pm$125}  & 36082\,{\scriptsize$\pm$2663} & 35589\,{\scriptsize$\pm$1884} \\
\rowcolor{gray!12}
& Avg. & 34875\,{\scriptsize$\pm$3118} & \textbf{33919}\,{\scriptsize$\pm$\textbf{3357}} & 35102\,{\scriptsize$\pm$3689} & 34674\,{\scriptsize$\pm$4070} & 34195\,{\scriptsize$\pm$3745} \\

\addlinespace[1pt]
\multirow{4}{*}{Eros. $\downarrow$}
& S1   & 0.0113\,{\scriptsize$\pm$0.0196} & 0.0948\,{\scriptsize$\pm$0.1179} & 0.1476\,{\scriptsize$\pm$0.0819} & 0.1406\,{\scriptsize$\pm$0.0670} & 0.0102\,{\scriptsize$\pm$0.0177} \\
& S2   & 0.0817\,{\scriptsize$\pm$0.0101} & 0.0842\,{\scriptsize$\pm$0.0189} & 0.1151\,{\scriptsize$\pm$0.0717} & 0.0613\,{\scriptsize$\pm$0.0253} & 0.0673\,{\scriptsize$\pm$0.0319} \\
& S3   & 0.2167\,{\scriptsize$\pm$0.0463} & 0.2264\,{\scriptsize$\pm$0.0578} & 0.2075\,{\scriptsize$\pm$0.0312} & 0.1730\,{\scriptsize$\pm$0.0358} & 0.2185\,{\scriptsize$\pm$0.0335} \\
\rowcolor{gray!12}
& Avg. & 0.1032\,{\scriptsize$\pm$0.0939} & 0.1352\,{\scriptsize$\pm$0.0954} & 0.1567\,{\scriptsize$\pm$0.0697} & 0.1250\,{\scriptsize$\pm$0.0639} & \textbf{0.0986}\,{\scriptsize$\pm$\textbf{0.0964}} \\

\addlinespace[1pt]
\multirow{4}{*}{Verb. $\downarrow$}
& S1   & 0.0670\,{\scriptsize$\pm$0.0189} & 0.0514\,{\scriptsize$\pm$0.0244} & 0.0148\,{\scriptsize$\pm$0.0056} & 0.0277\,{\scriptsize$\pm$0.0192} & 0.0361\,{\scriptsize$\pm$0.0251} \\
& S2   & 0.2261\,{\scriptsize$\pm$0.0336} & 0.2187\,{\scriptsize$\pm$0.0318} & 0.1986\,{\scriptsize$\pm$0.0610} & 0.1995\,{\scriptsize$\pm$0.0166} & 0.1382\,{\scriptsize$\pm$0.0212} \\
& S3   & 0.1878\,{\scriptsize$\pm$0.0234} & 0.1715\,{\scriptsize$\pm$0.0120} & 0.2090\,{\scriptsize$\pm$0.0335} & 0.1838\,{\scriptsize$\pm$0.0142} & 0.1238\,{\scriptsize$\pm$0.0127} \\
\rowcolor{gray!12}
& Avg. & 0.1603\,{\scriptsize$\pm$0.0754} & 0.1472\,{\scriptsize$\pm$0.0776} & 0.1408\,{\scriptsize$\pm$0.1008} & 0.1370\,{\scriptsize$\pm$0.0836} & \textbf{0.0994}\,{\scriptsize$\pm$\textbf{0.0510}} \\

\midrule
\multicolumn{7}{l}{\textbf{Library Size}} \\

\multirow{4}{*}{LOC}
& S1   & -- & 247\,{\scriptsize$\pm$41}  & 493\,{\scriptsize$\pm$59}  & 510\,{\scriptsize$\pm$62}  & 750\,{\scriptsize$\pm$151} \\
& S2   & -- & 225\,{\scriptsize$\pm$252} & 431\,{\scriptsize$\pm$116} & 499\,{\scriptsize$\pm$169} & 418\,{\scriptsize$\pm$179} \\
& S3   & -- & 194\,{\scriptsize$\pm$80}  & 418\,{\scriptsize$\pm$182} & 290\,{\scriptsize$\pm$67}  & 553\,{\scriptsize$\pm$123} \\
\rowcolor{gray!12}
& Avg. & -- & 222\,{\scriptsize$\pm$136} & 447\,{\scriptsize$\pm$117} & 433\,{\scriptsize$\pm$144} & 573\,{\scriptsize$\pm$196} \\

\addlinespace[1pt]
\multirow{4}{*}{Tok.}
& S1   & -- & 1849\,{\scriptsize$\pm$357}  & 3529\,{\scriptsize$\pm$399}  & 3666\,{\scriptsize$\pm$336} & 5281\,{\scriptsize$\pm$1103} \\
& S2   & -- & 1579\,{\scriptsize$\pm$1663} & 3123\,{\scriptsize$\pm$753}  & 3592\,{\scriptsize$\pm$750} & 3070\,{\scriptsize$\pm$1207} \\
& S3   & -- & 1334\,{\scriptsize$\pm$554}  & 2867\,{\scriptsize$\pm$1161} & 2012\,{\scriptsize$\pm$607} & 4077\,{\scriptsize$\pm$989} \\
\rowcolor{gray!12}
& Avg. & -- & 1587\,{\scriptsize$\pm$922}  & 3173\,{\scriptsize$\pm$776}  & 3090\,{\scriptsize$\pm$957} & 4143\,{\scriptsize$\pm$1353} \\

\addlinespace[1pt]
\multirow{4}{*}{MDL}
& S1   & -- & 640\,{\scriptsize$\pm$130} & 1297\,{\scriptsize$\pm$289} & 1445\,{\scriptsize$\pm$155} & 1985\,{\scriptsize$\pm$308} \\
& S2   & -- & 752\,{\scriptsize$\pm$753} & 1193\,{\scriptsize$\pm$299} & 1332\,{\scriptsize$\pm$286} & 1214\,{\scriptsize$\pm$383} \\
& S3   & -- & 526\,{\scriptsize$\pm$203} & 1147\,{\scriptsize$\pm$318} & 749\,{\scriptsize$\pm$141}  & 1693\,{\scriptsize$\pm$305} \\
\rowcolor{gray!12}
& Avg. & -- & 639\,{\scriptsize$\pm$407} & 1212\,{\scriptsize$\pm$270} & 1175\,{\scriptsize$\pm$369} & 1631\,{\scriptsize$\pm$444} \\

\bottomrule
\end{tabular}%
}
\caption{\textbf{Per-suite results on WebGen-Bench.} We report functionality and maintainability metrics across three disjoint suites of eight tasks each, as mean $\pm$ std over three independent trials per suite. Avg. is pooled across all 9 trials (3 suites $\times$ 3 trials).}
\label{tab:webgenbench-suite-results}
\end{table*}

\begin{table*}[t]
\centering
\scriptsize
\setlength{\tabcolsep}{4pt}
\renewcommand{\arraystretch}{1.05}
\resizebox{0.8\textwidth}{!}{%
\begin{tabular}{llccccc}
\toprule
\textbf{Metric}
& \textbf{Suite}
& \textbf{Zero}
& \textbf{Librarian}
& \multicolumn{2}{c}{\textbf{Naive}}
& \textbf{Full} \\
\cmidrule(lr){5-6}
&
&
&
& {\tiny \textbf{Implicit}}
& {\tiny \textbf{Ward}}
& \\
\midrule

\multicolumn{7}{l}{\textbf{Functionality}} \\

\multirow{6}{*}{Score $\uparrow$}
& S1   & 0.3409\,{\scriptsize$\pm$0.0918} & 0.3539\,{\scriptsize$\pm$0.0476} & 0.3393\,{\scriptsize$\pm$0.0822} & 0.3918\,{\scriptsize$\pm$0.0594} & 0.3690\,{\scriptsize$\pm$0.0456} \\
& S2   & 0.4394\,{\scriptsize$\pm$0.0127} & 0.3924\,{\scriptsize$\pm$0.0137} & 0.4885\,{\scriptsize$\pm$0.0382} & 0.4557\,{\scriptsize$\pm$0.0269} & 0.4584\,{\scriptsize$\pm$0.0104} \\
& S3   & 0.4461\,{\scriptsize$\pm$0.0193} & 0.4420\,{\scriptsize$\pm$0.0406} & 0.4006\,{\scriptsize$\pm$0.0154} & 0.3958\,{\scriptsize$\pm$0.0799} & 0.4160\,{\scriptsize$\pm$0.0597} \\
& S4   & 0.5909\,{\scriptsize$\pm$0.0564} & 0.5801\,{\scriptsize$\pm$0.0772} & 0.6071\,{\scriptsize$\pm$0.0701} & 0.5392\,{\scriptsize$\pm$0.0650} & 0.5751\,{\scriptsize$\pm$0.0683} \\
& S5   & 0.5261\,{\scriptsize$\pm$0.0674} & 0.5271\,{\scriptsize$\pm$0.0206} & 0.5655\,{\scriptsize$\pm$0.0440} & 0.5773\,{\scriptsize$\pm$0.0654} & 0.5860\,{\scriptsize$\pm$0.0174} \\
\rowcolor{gray!12}
& Avg. & 0.4687\,{\scriptsize$\pm$0.1004} & 0.4591\,{\scriptsize$\pm$0.0949} & 0.4802\,{\scriptsize$\pm$0.1132} & 0.4720\,{\scriptsize$\pm$0.0936} & \textbf{0.4809}\,{\scriptsize$\pm$\textbf{0.0975}} \\

\midrule
\multicolumn{7}{l}{\textbf{Maintainability}} \\

\multirow{6}{*}{LOC $\downarrow$}
& S1   & 7128\,{\scriptsize$\pm$891}  & 7125\,{\scriptsize$\pm$874}  & 7155\,{\scriptsize$\pm$582}  & 7156\,{\scriptsize$\pm$602} & 7080\,{\scriptsize$\pm$526} \\
& S2   & 6239\,{\scriptsize$\pm$1006} & 6209\,{\scriptsize$\pm$961}  & 6288\,{\scriptsize$\pm$754}  & 6381\,{\scriptsize$\pm$820} & 6387\,{\scriptsize$\pm$1109} \\
& S3   & 6922\,{\scriptsize$\pm$86}   & 6928\,{\scriptsize$\pm$120}  & 6309\,{\scriptsize$\pm$518}  & 5992\,{\scriptsize$\pm$337} & 6198\,{\scriptsize$\pm$381} \\
& S4   & 6057\,{\scriptsize$\pm$421}  & 5998\,{\scriptsize$\pm$407} & 6041\,{\scriptsize$\pm$455}  & 6134\,{\scriptsize$\pm$1059} & 5604\,{\scriptsize$\pm$528} \\
& S5   & 6544\,{\scriptsize$\pm$488}  & 6496\,{\scriptsize$\pm$437}  & 6430\,{\scriptsize$\pm$140}  & 6536\,{\scriptsize$\pm$286} & 5993\,{\scriptsize$\pm$441} \\
\rowcolor{gray!12}
& Avg. & 6578\,{\scriptsize$\pm$701}  & 6551\,{\scriptsize$\pm$697}  & 6445\,{\scriptsize$\pm$594}  & 6440\,{\scriptsize$\pm$715} & \textbf{6252}\,{\scriptsize$\pm$\textbf{748}} \\

\addlinespace[1pt]
\multirow{6}{*}{Tok. $\downarrow$}
& S1   & 69723\,{\scriptsize$\pm$6679}  & 69730\,{\scriptsize$\pm$6560} & 70292\,{\scriptsize$\pm$4967} & 70942\,{\scriptsize$\pm$5114} & 67897\,{\scriptsize$\pm$5531} \\
& S2   & 64270\,{\scriptsize$\pm$10973} & 64100\,{\scriptsize$\pm$10502} & 65198\,{\scriptsize$\pm$6768} & 64823\,{\scriptsize$\pm$9037} & 63029\,{\scriptsize$\pm$12116} \\
& S3   & 73351\,{\scriptsize$\pm$2528}  & 73147\,{\scriptsize$\pm$1806} & 64868\,{\scriptsize$\pm$6821} & 59797\,{\scriptsize$\pm$4782} & 62997\,{\scriptsize$\pm$5168} \\
& S4   & 64699\,{\scriptsize$\pm$4074}  & 64160\,{\scriptsize$\pm$3727} & 63268\,{\scriptsize$\pm$6188} & 64403\,{\scriptsize$\pm$10701} & 59037\,{\scriptsize$\pm$4488} \\
& S5   & 70974\,{\scriptsize$\pm$7082}  & 70350\,{\scriptsize$\pm$6461} & 69066\,{\scriptsize$\pm$1807} & 70176\,{\scriptsize$\pm$2483} & 64613\,{\scriptsize$\pm$4752} \\
\rowcolor{gray!12}
& Avg. & 68603\,{\scriptsize$\pm$6900}  & 68297\,{\scriptsize$\pm$6644} & 66539\,{\scriptsize$\pm$5506} & 66028\,{\scriptsize$\pm$7346} & \textbf{63514}\,{\scriptsize$\pm$\textbf{6636}} \\

\addlinespace[1pt]
\multirow{6}{*}{MDL $\downarrow$}
& S1   & 28483\,{\scriptsize$\pm$2831} & 28724\,{\scriptsize$\pm$2727} & 29080\,{\scriptsize$\pm$2283} & 28671\,{\scriptsize$\pm$2073} & 29006\,{\scriptsize$\pm$3820} \\
& S2   & 29510\,{\scriptsize$\pm$4585} & 29656\,{\scriptsize$\pm$4605} & 29216\,{\scriptsize$\pm$1649} & 30483\,{\scriptsize$\pm$4302} & 30444\,{\scriptsize$\pm$7164} \\
& S3   & 36079\,{\scriptsize$\pm$3979} & 36140\,{\scriptsize$\pm$3409} & 32919\,{\scriptsize$\pm$2949} & 29731\,{\scriptsize$\pm$2625} & 32518\,{\scriptsize$\pm$624} \\
& S4   & 27663\,{\scriptsize$\pm$1995} & 27541\,{\scriptsize$\pm$1892} & 27665\,{\scriptsize$\pm$4704} & 28750\,{\scriptsize$\pm$4919} & 27281\,{\scriptsize$\pm$1832} \\
& S5   & 29977\,{\scriptsize$\pm$3606} & 29941\,{\scriptsize$\pm$3555} & 30369\,{\scriptsize$\pm$1720} & 30207\,{\scriptsize$\pm$3485} & 28198\,{\scriptsize$\pm$1530} \\
\rowcolor{gray!12}
& Avg. & 30342\,{\scriptsize$\pm$4283} & 30400\,{\scriptsize$\pm$4201} & 29850\,{\scriptsize$\pm$3045} & 29568\,{\scriptsize$\pm$3166} & \textbf{29489}\,{\scriptsize$\pm$\textbf{3728}} \\

\addlinespace[1pt]
\multirow{6}{*}{Eros. $\downarrow$}
& S1   & 0.2070\,{\scriptsize$\pm$0.0202} & 0.2133\,{\scriptsize$\pm$0.0290} & 0.2536\,{\scriptsize$\pm$0.0835} & 0.2585\,{\scriptsize$\pm$0.0159} & 0.1552\,{\scriptsize$\pm$0.0244} \\
& S2   & 0.2342\,{\scriptsize$\pm$0.0777} & 0.2350\,{\scriptsize$\pm$0.0784} & 0.2394\,{\scriptsize$\pm$0.0460} & 0.2281\,{\scriptsize$\pm$0.0769} & 0.2678\,{\scriptsize$\pm$0.0348} \\
& S3   & 0.2560\,{\scriptsize$\pm$0.0660} & 0.2383\,{\scriptsize$\pm$0.0519} & 0.2656\,{\scriptsize$\pm$0.0509} & 0.2709\,{\scriptsize$\pm$0.0995} & 0.2616\,{\scriptsize$\pm$0.0347} \\
& S4   & 0.2012\,{\scriptsize$\pm$0.1213} & 0.1974\,{\scriptsize$\pm$0.1132} & 0.2978\,{\scriptsize$\pm$0.0476} & 0.2916\,{\scriptsize$\pm$0.0856} & 0.1641\,{\scriptsize$\pm$0.0555} \\
& S5   & 0.3804\,{\scriptsize$\pm$0.0646} & 0.3795\,{\scriptsize$\pm$0.0657} & 0.3849\,{\scriptsize$\pm$0.1232} & 0.3314\,{\scriptsize$\pm$0.0611} & 0.2967\,{\scriptsize$\pm$0.1291} \\
\rowcolor{gray!12}
& Avg. & 0.2558\,{\scriptsize$\pm$0.0939} & 0.2527\,{\scriptsize$\pm$0.0915} & 0.2882\,{\scriptsize$\pm$0.0840} & 0.2761\,{\scriptsize$\pm$0.0717} & \textbf{0.2291}\,{\scriptsize$\pm$\textbf{0.0828}} \\

\addlinespace[1pt]
\multirow{6}{*}{Verb. $\downarrow$}
& S1   & 0.8889\,{\scriptsize$\pm$0.0042} & 0.8865\,{\scriptsize$\pm$0.0040} & 0.8707\,{\scriptsize$\pm$0.0216} & 0.8860\,{\scriptsize$\pm$0.0135} & 0.8552\,{\scriptsize$\pm$0.0064} \\
& S2   & 0.8482\,{\scriptsize$\pm$0.0231} & 0.8435\,{\scriptsize$\pm$0.0258} & 0.8413\,{\scriptsize$\pm$0.0197} & 0.8582\,{\scriptsize$\pm$0.0102} & 0.8356\,{\scriptsize$\pm$0.0154} \\
& S3   & 0.8455\,{\scriptsize$\pm$0.0102} & 0.8404\,{\scriptsize$\pm$0.0120} & 0.8394\,{\scriptsize$\pm$0.0217} & 0.8404\,{\scriptsize$\pm$0.0278} & 0.8393\,{\scriptsize$\pm$0.0221} \\
& S4   & 0.8645\,{\scriptsize$\pm$0.0097} & 0.8618\,{\scriptsize$\pm$0.0117} & 0.8719\,{\scriptsize$\pm$0.0090} & 0.8646\,{\scriptsize$\pm$0.0048} & 0.8518\,{\scriptsize$\pm$0.0222} \\
& S5   & 0.8853\,{\scriptsize$\pm$0.0153} & 0.8818\,{\scriptsize$\pm$0.0173} & 0.8675\,{\scriptsize$\pm$0.0112} & 0.8734\,{\scriptsize$\pm$0.0251} & 0.8630\,{\scriptsize$\pm$0.0132} \\
\rowcolor{gray!12}
& Avg. & 0.8665\,{\scriptsize$\pm$0.0221} & 0.8628\,{\scriptsize$\pm$0.0238} & 0.8582\,{\scriptsize$\pm$0.0212} & 0.8645\,{\scriptsize$\pm$0.0222} & \textbf{0.8490}\,{\scriptsize$\pm$\textbf{0.0177}} \\

\midrule
\multicolumn{7}{l}{\textbf{Library Size}} \\

\multirow{6}{*}{LOC}
& S1   & -- & 152\,{\scriptsize$\pm$53} & 216\,{\scriptsize$\pm$91}  & 308\,{\scriptsize$\pm$83}  & 541\,{\scriptsize$\pm$73} \\
& S2   & -- & 48\,{\scriptsize$\pm$37}  & 156\,{\scriptsize$\pm$87}  & 355\,{\scriptsize$\pm$109} & 277\,{\scriptsize$\pm$150} \\
& S3   & -- & 105\,{\scriptsize$\pm$48} & 220\,{\scriptsize$\pm$49}  & 231\,{\scriptsize$\pm$167} & 217\,{\scriptsize$\pm$128} \\
& S4   & -- & 163\,{\scriptsize$\pm$43} & 320\,{\scriptsize$\pm$25}  & 364\,{\scriptsize$\pm$79}  & 524\,{\scriptsize$\pm$15} \\
& S5   & -- & 20\,{\scriptsize$\pm$5}   & 146\,{\scriptsize$\pm$80}  & 215\,{\scriptsize$\pm$70}  & 118\,{\scriptsize$\pm$34} \\
\rowcolor{gray!12}
& Avg. & -- & 98\,{\scriptsize$\pm$68}  & 212\,{\scriptsize$\pm$88}  & 295\,{\scriptsize$\pm$111} & 336\,{\scriptsize$\pm$192} \\

\addlinespace[1pt]
\multirow{6}{*}{Tok.}
& S1   & -- & 1445\,{\scriptsize$\pm$501} & 1960\,{\scriptsize$\pm$821} & 2914\,{\scriptsize$\pm$712}  & 4872\,{\scriptsize$\pm$754} \\
& S2   & -- & 472\,{\scriptsize$\pm$364}  & 1414\,{\scriptsize$\pm$629} & 3132\,{\scriptsize$\pm$963}  & 2733\,{\scriptsize$\pm$1786} \\
& S3   & -- & 950\,{\scriptsize$\pm$397}  & 2104\,{\scriptsize$\pm$587} & 2089\,{\scriptsize$\pm$1418} & 1879\,{\scriptsize$\pm$1135} \\
& S4   & -- & 2017\,{\scriptsize$\pm$346} & 3385\,{\scriptsize$\pm$401} & 4048\,{\scriptsize$\pm$899}  & 5309\,{\scriptsize$\pm$184} \\
& S5   & -- & 171\,{\scriptsize$\pm$48}   & 1347\,{\scriptsize$\pm$653} & 1757\,{\scriptsize$\pm$415}  & 1050\,{\scriptsize$\pm$416} \\
\rowcolor{gray!12}
& Avg. & -- & 1011\,{\scriptsize$\pm$752} & 2042\,{\scriptsize$\pm$929} & 2788\,{\scriptsize$\pm$1155} & 3168\,{\scriptsize$\pm$1927} \\

\addlinespace[1pt]
\multirow{6}{*}{MDL}
& S1   & -- & 516\,{\scriptsize$\pm$208} & 650\,{\scriptsize$\pm$190} & 901\,{\scriptsize$\pm$125}  & 1583\,{\scriptsize$\pm$263} \\
& S2   & -- & 201\,{\scriptsize$\pm$126} & 511\,{\scriptsize$\pm$218} & 1212\,{\scriptsize$\pm$381} & 1043\,{\scriptsize$\pm$441} \\
& S3   & -- & 368\,{\scriptsize$\pm$130} & 854\,{\scriptsize$\pm$131} & 825\,{\scriptsize$\pm$561}  & 786\,{\scriptsize$\pm$379} \\
& S4   & -- & 779\,{\scriptsize$\pm$240} & 1137\,{\scriptsize$\pm$148} & 1300\,{\scriptsize$\pm$149} & 2084\,{\scriptsize$\pm$125} \\
& S5   & -- & 121\,{\scriptsize$\pm$26}  & 673\,{\scriptsize$\pm$327} & 896\,{\scriptsize$\pm$41}   & 520\,{\scriptsize$\pm$197} \\
\rowcolor{gray!12}
& Avg. & -- & 397\,{\scriptsize$\pm$280} & 765\,{\scriptsize$\pm$287} & 1027\,{\scriptsize$\pm$332} & 1203\,{\scriptsize$\pm$637} \\

\bottomrule
\end{tabular}%
}
\caption{\textbf{Per-suite results on PaperBench.} We report functionality and maintainability metrics across five disjoint suites of four tasks each, as mean $\pm$ std over three independent trials per suite. Avg. is pooled across all 15 trials (5 suites $\times$ 3 trials). Library Size reports the size of the final Super Library only.}
\label{tab:paperbench-suite-results}
\end{table*}

\subsection{Qualitative Example}
\label{sec:appendix:initial-example}

\begin{figure*}[t]
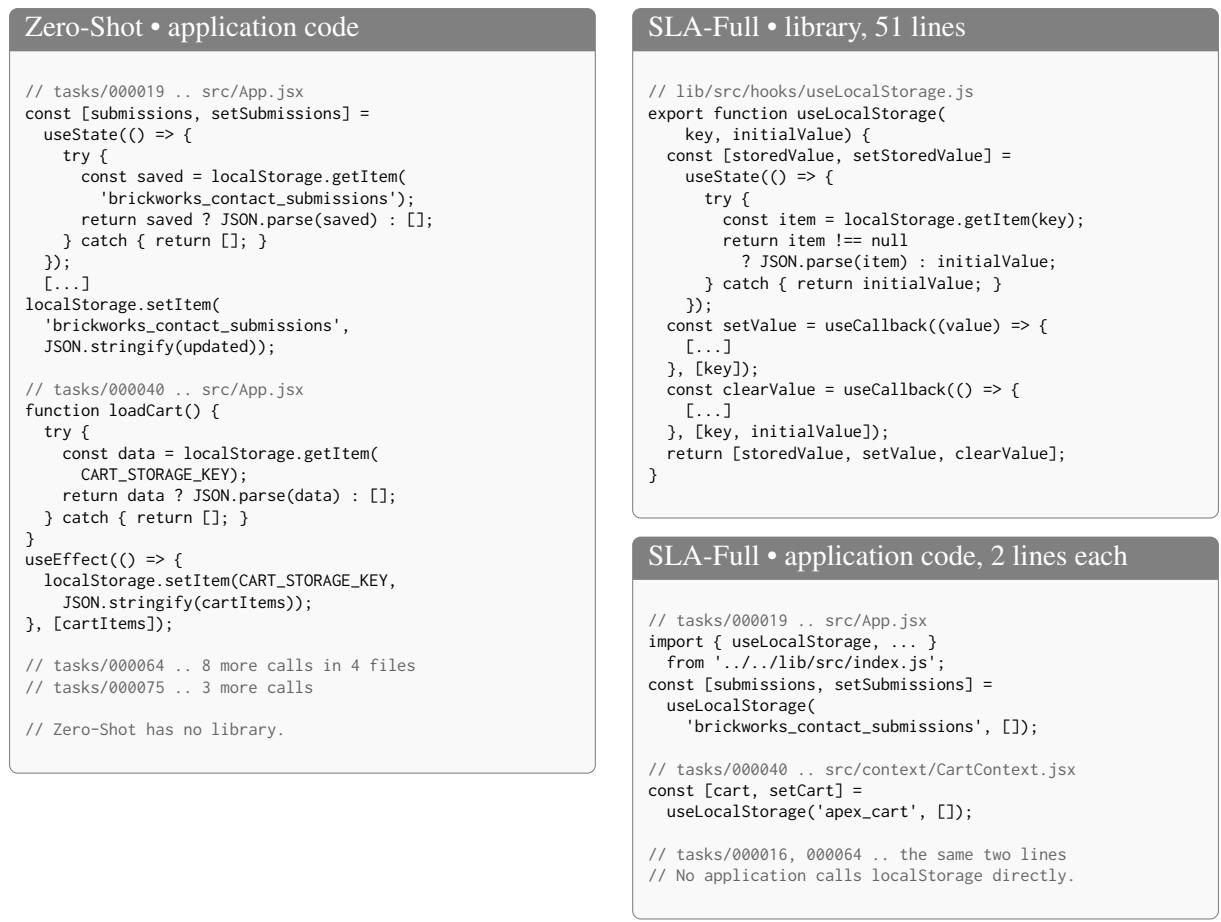

\begin{minipage}[t]{0.485\textwidth}
\vspace{0pt}
\begin{codebox}[title={Zero-Shot \textbullet{} application code}]
// tasks/000019 .. src/App.jsx
const [submissions, setSubmissions] =
  useState(() => {
    try {
      const saved = localStorage.getItem(
        'brickworks_contact_submissions');
      return saved ? JSON.parse(saved) : [];
    } catch { return []; }
  });
  [...]
localStorage.setItem(
  'brickworks_contact_submissions',
  JSON.stringify(updated));

// tasks/000040 .. src/App.jsx
function loadCart() {
  try {
    const data = localStorage.getItem(
      CART_STORAGE_KEY);
    return data ? JSON.parse(data) : [];
  } catch { return []; }
}
useEffect(() => {
  localStorage.setItem(CART_STORAGE_KEY,
    JSON.stringify(cartItems));
}, [cartItems]);

// tasks/000064 .. 8 more calls in 4 files
// tasks/000075 .. 3 more calls

// Zero-Shot has no library.
\end{codebox}
\end{minipage}\hfill
\begin{minipage}[t]{0.485\textwidth}
\vspace{0pt}
\begin{codebox}[title={SLA-Full \textbullet{} library, 51 lines}]
// lib/src/hooks/useLocalStorage.js
export function useLocalStorage(
    key, initialValue) {
  const [storedValue, setStoredValue] =
    useState(() => {
      try {
        const item = localStorage.getItem(key);
        return item !== null
          ? JSON.parse(item) : initialValue;
      } catch { return initialValue; }
    });
  const setValue = useCallback((value) => {
    [...]
  }, [key]);
  const clearValue = useCallback(() => {
    [...]
  }, [key, initialValue]);
  return [storedValue, setValue, clearValue];
}
\end{codebox}

\begin{codebox}[title={SLA-Full \textbullet{} application code, 2 lines each}]
// tasks/000019 .. src/App.jsx
import { useLocalStorage, ... }
  from '../../lib/src/index.js';
const [submissions, setSubmissions] =
  useLocalStorage(
    'brickworks_contact_submissions', []);

// tasks/000040 .. src/context/CartContext.jsx
const [cart, setCart] =
  useLocalStorage('apex_cart', []);

// tasks/000016, 000064 .. the same two lines
// No application calls localStorage directly.
\end{codebox}
\end{minipage}
\caption{\textbf{Initial-construction example.} Zero-Shot repeats the same
persistence cycle in every application that needs it, while \textsc{SLA-Full}
defines it once in the library.}
\label{fig:construction-example}
\end{figure*}

Figure~\ref{fig:construction-example} shows a detailed example from
WebGen-Bench Suite~1. Four of the eight applications persist state to
\texttt{localStorage} in both portfolios.
Zero-Shot writes the same load-and-persist cycle into each of them, each with its
own \texttt{try}/\texttt{catch} and JSON round-trip. \textsc{SLA-Full} extracts
the cycle into one library hook, which the four applications reach through one
import and one call each.

\section{Post-Construction Maintenance}
\label{sec:appendix:maintenance}

\subsection{Policy Updates}
\label{sec:appendix:maintenance-inputs}

We construct each maintenance request in two isolated stages. First, an agent observes only the task metadata for a suite and writes one policy update grounded in behavior shared across its applications. Second, isolated sessions use each application's original instruction and layout description to instantiate the policy for that application. Neither stage observes generated code, Super Libraries, or evaluation results. We retain the first policy produced for each suite without selecting among alternatives. This appendix reproduces the suite-level instructions and the resulting change-specific WebVoyager tests. The tests also identify every target application.

\paragraph{Suite 1: Engagement-form standardization.}
\begin{quote}
\small
Our organization has standardized how all customer-facing engagement forms behave across its web properties. For every contact, lead, newsletter, reservation, or similar primary engagement form:
\begin{enumerate}[leftmargin=*,nosep]
    \item Required text inputs must reject empty or whitespace-only values.
    \item If an email field exists, it must require a value containing both ``@'' and ``.''.
    \item An invalid submit must show a clear inline error message and must not show a success state.
    \item An invalid submit must preserve the user's current input values.
    \item A valid submit must show an in-page confirmation without navigating away.
    \item The confirmation must include either ``Received'' or ``Submitted'', and must include the submitted name or email when such a field exists.
\end{enumerate}
Original navigation and content behavior must not be removed or weakened.
\end{quote}

\paragraph{Suite 2: Record-audit policy.}
\begin{quote}
\small
Our organization has adopted a record-audit policy for its data-management applications. For each app's primary create/update workflows:
\begin{enumerate}[leftmargin=*,nosep]
    \item When a new domain record is created, display a human-readable record ID with a domain prefix and a 3+ digit/character suffix where possible (e.g., CALL-001, LEAD-001, PAT-001, TASK-001). Where the app already exposes a domain-prefixed ID scheme, keep that scheme.
    \item Created or updated records must display a ``Last updated'' timestamp or date, refreshed on update.
    \item Create and update actions must show a confirmation message.
    \item The created/updated record must be findable from the app's relevant list, search, filter, tracking, or status view.
\end{enumerate}
Original CRUD, report, and navigation behavior must not be removed or weakened.
\end{quote}

\paragraph{Suite 3: Trust, safety, and search quality.}
\begin{quote}
\small
Our platform group has adopted a trust-and-safety policy for user-generated content, plus a search-quality baseline.

\textbf{Part A: Moderation status.} For each publicly visible user-generated content submission flow---submissions that appear in lists, feeds, or search results seen by other users (e.g., shared promotions, job postings, job/internship applications appearing in a reviewable list, tool comments/ratings, uploaded pictures, newly registered public profile cards):
\begin{enumerate}[leftmargin=*,nosep]
    \item After successful submission, show a confirmation message.
    \item The submitted item must display a visible ``Pending review'' badge or status text.
    \item Pending items must remain visible in the relevant list/detail/search views---do not hide them.
\end{enumerate}
Explicitly excluded from moderation (keep original immediate behavior): edits to one's own profile fields, one-to-one chat messages, friend invitations.

\textbf{Part B: Search normalization.} Search must trim leading/trailing whitespace, match case-insensitively, and show a clear empty-state message when there are no matches (instead of a blank area or silently showing all items).

Original registration, login, navigation, and search behavior must not be removed or weakened.
\end{quote}

\subsection{Expanded Results}
\label{sec:appendix:maintenance-results}

Table~\ref{tab:maintenance-full} supplements the aggregate results in Table~\ref{tab:maintenance} with the number of files touched, changes in portfolio-level structural metrics, and Patch Size for each suite.

\subsection{Qualitative Example}
\label{sec:appendix:maintenance-example}

\begin{figure*}[t]
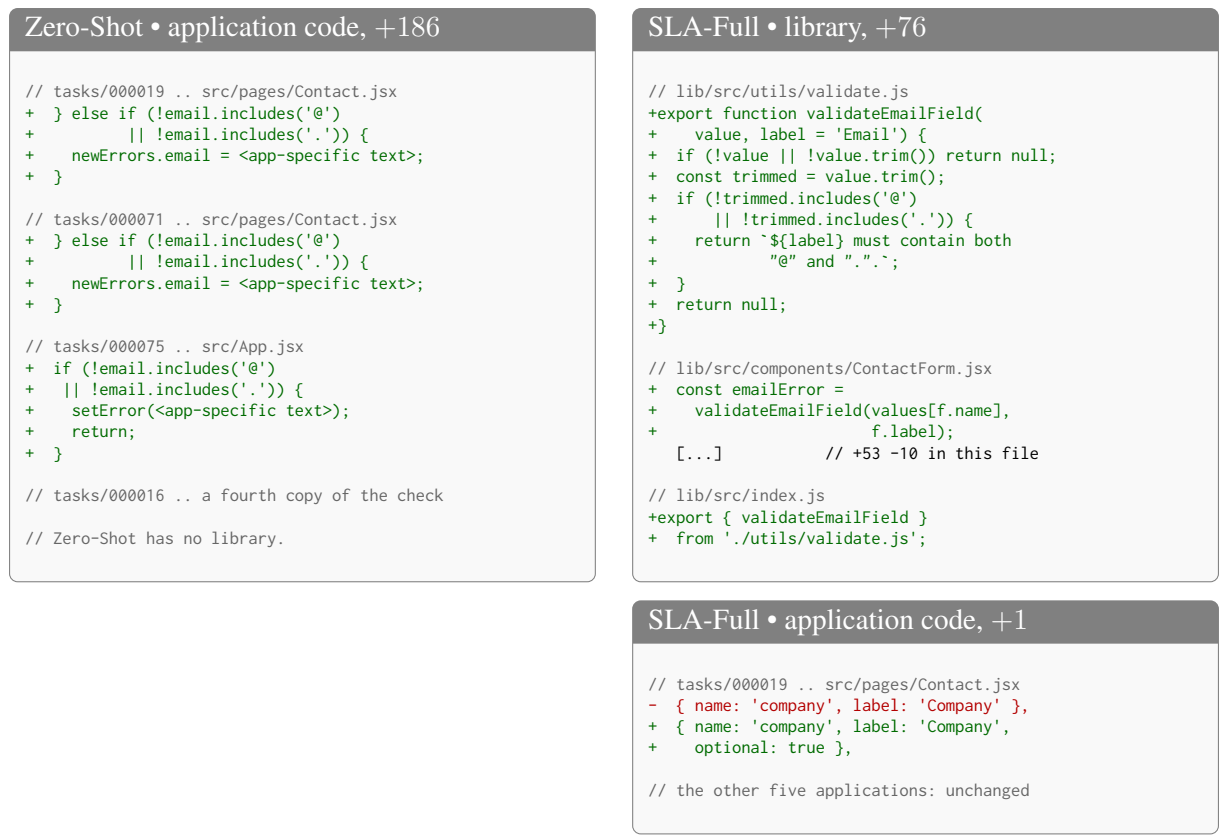

\begin{minipage}[t]{0.485\textwidth}
\vspace{0pt}
\begin{codebox}[title={Zero-Shot \textbullet{} application code, $+186$}]
// tasks/000019 .. src/pages/Contact.jsx
+  } else if (!email.includes('@')
+          || !email.includes('.')) {
+    newErrors.email = <app-specific text>;
+  }

// tasks/000071 .. src/pages/Contact.jsx
+  } else if (!email.includes('@')
+          || !email.includes('.')) {
+    newErrors.email = <app-specific text>;
+  }

// tasks/000075 .. src/App.jsx
+  if (!email.includes('@')
+   || !email.includes('.')) {
+    setError(<app-specific text>);
+    return;
+  }

// tasks/000016 .. a fourth copy of the check

// Zero-Shot has no library.
\end{codebox}
\end{minipage}\hfill
\begin{minipage}[t]{0.485\textwidth}
\vspace{0pt}
\begin{codebox}[title={SLA-Full \textbullet{} library, $+76$}]
// lib/src/utils/validate.js
+export function validateEmailField(
+    value, label = 'Email') {
+  if (!value || !value.trim()) return null;
+  const trimmed = value.trim();
+  if (!trimmed.includes('@')
+      || !trimmed.includes('.')) {
+    return `${label} must contain both
+            "@" and ".".`;
+  }
+  return null;
+}

// lib/src/components/ContactForm.jsx
+  const emailError =
+    validateEmailField(values[f.name],
+                       f.label);
   [...]           // +53 -10 in this file

// lib/src/index.js
+export { validateEmailField }
+  from './utils/validate.js';
\end{codebox}

\begin{codebox}[title={SLA-Full \textbullet{} application code, $+1$}]
// tasks/000019 .. src/pages/Contact.jsx
-  { name: 'company', label: 'Company' },
+  { name: 'company', label: 'Company',
+    optional: true },

// the other five applications: unchanged
\end{codebox}
\end{minipage}
\caption{\textbf{Post-construction maintenance example.} Zero-Shot repeats the
requested rule in every application it applies to, while \textsc{SLA-Full} adds
it once to the library.}
\label{fig:maintenance-example}
\end{figure*}

Figure~\ref{fig:maintenance-example} shows a detailed example from
WebGen-Bench Suite~1. Zero-Shot and \textsc{SLA-Full} receive the same
policy (Section~\ref{sec:appendix:maintenance-inputs}), whose first two items
require that an email field contain both \texttt{@} and a dot. Zero-Shot
implements the rule independently in four of the six applications, each with its
own message. \textsc{SLA-Full} adds it once to a library component, the shared
validator that the shared contact form already calls, so five of the six
applications need no edit at all.

\subsection{Change-Specific Test Cases}
\label{sec:appendix:maintenance-tests}

Tables~\ref{tab:maintenance-tests-suite1}--\ref{tab:maintenance-tests-suite3} report the WebVoyager action and expected result used to evaluate each target application.

\begin{table*}[t]
\centering
\small
\setlength{\tabcolsep}{5pt}
\renewcommand{\arraystretch}{1.08}
\resizebox{0.92\textwidth}{!}{%
\begin{tabular}{lcccccccccccc}
\toprule
& \multicolumn{4}{c}{\textbf{Patch}}
& \multicolumn{5}{c}{\textbf{Change in maintainability}}
& \multicolumn{3}{c}{\textbf{Patch Size by suite}} \\
\cmidrule(lr){2-5} \cmidrule(lr){6-10} \cmidrule(lr){11-13}
\textbf{Method}
& Added LOC $\downarrow$ & App $\downarrow$ & Lib & Files $\downarrow$
& $\Delta$LOC $\downarrow$ & $\Delta$Tok $\downarrow$ & $\Delta$MDL $\downarrow$ & $\Delta$Eros. $\downarrow$ & $\Delta$Verb. $\downarrow$
& S1 & S2 & S3 \\
\midrule
Zero-Shot
& 936 & 936 & 0 & 29.8
& $+$605 & $+$5101 & $+$2565 & $-$0.004 & $+$0.007
& 182 & 1804 & 822 \\
\textsc{Librarian}
& 522 & 500 & 22 & 27.0
& -- & -- & -- & -- & --
& 151 & 717 & 699 \\
\midrule
\textsc{Naive-Implicit}
& 632 & 618 & 14 & 27.4
& $+$278 & $+$2874 & $+$2078 & $+$0.019 & $+$0.005
& 97 & 1328 & 471 \\
\textsc{Naive-Ward}
& 380 & 363 & 18 & 22.2
& $+$251 & $+$2779 & $+$1909 & $+$0.030 & $+$0.011
& 103 & 574 & 463 \\
\rowcolor{gray!12}
\textsc{SLA-Full}
& \textbf{256} & \textbf{232} & 24 & \textbf{19.9}
& \textbf{$+$171} & \textbf{$+$2073} & \textbf{$+$1377} & $+$0.001 & \textbf{$+$0.003}
& \textbf{79} & \textbf{422} & \textbf{268} \\
\bottomrule
\end{tabular}%
}
\caption{\textbf{Post-construction maintenance, full results.} The table extends Table~\ref{tab:maintenance} with the number of files touched, changes in portfolio-level structural metrics, and Patch Size for each suite. $\Delta$ columns report post-patch minus pre-patch values. They are unavailable for \textsc{Librarian}, whose pre-patch portfolio is a post-hoc refactoring of the Zero-Shot codebases rather than a round-structured run.}
\label{tab:maintenance-full}
\end{table*}

\begin{table*}[p]
\centering
\scriptsize
\setlength{\tabcolsep}{3pt}
\renewcommand{\arraystretch}{1.03}
\caption{\textbf{Change-specific tests for Suite 1.} Each row reproduces one WebVoyager action and its expected result.}
\label{tab:maintenance-tests-suite1}
\begin{tabular}{p{0.08\textwidth}p{0.42\textwidth}p{0.43\textwidth}}
\toprule
\textbf{App ID} & \textbf{Action} & \textbf{Expected result} \\
\midrule
\texttt{000015} & Go to the Contact Us page. In the contact form, fill the Name and Message fields but leave Email empty (or whitespace only), then submit. & An inline error message appears, no success/confirmation state is shown, and the values already typed in Name and Message remain in the fields. \\
\texttt{000015} & On the same contact form, fill Name, a valid Email containing ``@'' and ``.'', and Message, then submit. & An in-page confirmation appears on the Contact Us page without navigating away, containing the word ``Received'' or ``Submitted'' and including the submitted name or email. \\
\texttt{000016} & Go to the Contact page. In the contact form, enter a Name and Message but type an invalid email without ``@'' or ``.'' (e.g., ``testmail''), then submit. & A clear inline error appears for the email field, no confirmation banner is shown, and the typed Name, Email, and Message values remain in the inputs. \\
\texttt{000016} & Correct the email to a valid address containing ``@'' and ``.'', then submit the form. & An in-page confirmation banner appears on the Contact page without navigating away, containing ``Received'' or ``Submitted'' together with the submitted name or email. \\
\texttt{000019} & Go to the Contact page. Submit the contact form with the ``How can we help?'' field empty while Name and Email are filled. & An inline error message appears, no confirmation is shown, and all values typed so far, including Company if entered, are preserved in the form. \\
\texttt{000019} & Fill Name, a valid Email containing ``@'' and ``.'', and the ``How can we help?'' field, then submit. & An in-page confirmation appears on the Contact page without navigating away, including ``Received'' or ``Submitted'' plus the submitted name or email. \\
\texttt{000064} & Go to the Newsletter page. Enter an invalid email without ``@'' or ``.'' (e.g., ``newsfan'') into the subscription input and click Subscribe. & A clear inline error message appears, no success state is shown, and the typed value remains in the input. \\
\texttt{000064} & Replace it with a valid email containing ``@'' and ``.'', then click Subscribe. & An in-page confirmation appears on the same page without navigating away, containing ``Received'' or ``Submitted'' along with the submitted email address. \\
\texttt{000071} & Go to the Contact page's ``Get in Touch'' form. Submit it with the Message field empty (or spaces only) while Name and Email are filled. & A clear inline error message appears, no saved/confirmation line is shown, and the typed Name and Email values are preserved in the form. \\
\texttt{000071} & Fill all three fields with valid values (email containing ``@'' and ``.'') and submit. & An in-page confirmation appears beneath the form without navigating away, including ``Received'' or ``Submitted'' together with the submitted name or email. \\
\texttt{000075} & Scroll to the Contact Us section. Submit the contact form with all fields empty. & A clear inline error message appears within the Contact Us section, and no success state is shown. \\
\texttt{000075} & Fill name, a valid email containing ``@'' and ``.'', and a message, then submit. & An in-page confirmation appears within the Contact Us section without navigating away, and includes ``Received'' or ``Submitted'' plus the submitted name or email. \\
\bottomrule
\end{tabular}
\end{table*}

\begin{table*}[p]
\centering
\scriptsize
\setlength{\tabcolsep}{3pt}
\renewcommand{\arraystretch}{1.03}
\caption{\textbf{Change-specific tests for Suite 2.} Each row reproduces one WebVoyager action and its expected result.}
\label{tab:maintenance-tests-suite2}
\begin{tabular}{p{0.08\textwidth}p{0.42\textwidth}p{0.43\textwidth}}
\toprule
\textbf{App ID} & \textbf{Action} & \textbf{Expected result} \\
\midrule
\texttt{000022} & Create a new call log via the New Call form with required fields filled. & A confirmation message appears, and the new call log displays a human-readable ID with the CALL- prefix (e.g., CALL-001) and a ``Last updated'' timestamp. The call appears in the Open Calls or To-Do Calls view. \\
\texttt{000022} & Modify an existing client's information (or close/mark done an existing call). & A confirmation message appears and the record's ``Last updated'' timestamp refreshes to reflect the change. \\
\texttt{000023} & Create a new lead via the New Lead form/modal with valid details. & A confirmation message appears, and the new lead is shown with a LEAD- prefixed ID (e.g., LEAD-001) and a ``Last updated'' timestamp or date, visible in the leads table. \\
\texttt{000023} & Change that lead's status (or edit it), then locate it using the leads search or filters. & A confirmation appears, the ``Last updated'' value refreshes, and the lead is findable via search/filter with its LEAD- ID still displayed. \\
\texttt{000044} & Register a new product via the registration form. & A confirmation message appears and the product's SCT- prefixed ID is displayed immediately upon registration. \\
\texttt{000044} & Add a checkpoint to that product on the Track screen, then look it up on the Inquiry screen. & A confirmation appears for the checkpoint, the product's ``Last updated'' line refreshes, and the product is findable via the Inquiry search with its updated status. \\
\texttt{000046} & Create a new patient electronic file in the patient management section. & A confirmation message appears and the new patient record displays a PAT- prefixed ID (e.g., PAT-001) and a ``Last updated'' timestamp or date, visible in the patient files list. \\
\texttt{000046} & Add a new medicine in the pharmacy section (or collect a patient fee in finances). & A confirmation appears and the new record shows its prefixed ID (MED- or FEE-) with a ``Last updated'' value, findable in the corresponding section list. \\
\texttt{000080} & Add a new task via the Add Task input. & A confirmation message appears and the new task row displays a TASK- prefixed ID (e.g., TASK-001) and a ``Last updated'' timestamp or date. \\
\texttt{000080} & Edit that task's description (or toggle it complete), then find it via search and the matching filter. & A confirmation appears, the ``Last updated'' value refreshes, the ID persists, and the task is findable via search and under the correct All/Active/Completed filter. \\
\texttt{000083} & Submit a real-time report on a case. & A confirmation message appears and the new report entry displays an RPT- prefixed ID with a timestamp, visible in the selected case's report history. \\
\texttt{000083} & Log work hours for the same case. & A confirmation appears, the new hours entry shows an HRS- prefixed ID with a ``Last updated'' timestamp, and the case's total hours update. \\
\texttt{000085} & Create a new project via the New Project action. & A confirmation message appears and the new project row displays a PRJ- prefixed ID (e.g., PRJ-001) and a ``Last updated'' date, visible in the Projects table and reflected on the dashboard. \\
\texttt{000085} & Add a schedule task (or a document), then mark the task complete. & A confirmation appears on create, and the record shows its prefixed ID (TASK- or DOC-) with a ``Last updated'' value that refreshes on the completion toggle. \\
\bottomrule
\end{tabular}
\end{table*}

\begin{table*}[p]
\centering
\scriptsize
\setlength{\tabcolsep}{3pt}
\renewcommand{\arraystretch}{1.03}
\caption{\textbf{Change-specific tests for Suite 3.} Each row reproduces one WebVoyager action and its expected result.}
\label{tab:maintenance-tests-suite3}
\begin{tabular}{p{0.08\textwidth}p{0.42\textwidth}p{0.43\textwidth}}
\toprule
\textbf{App ID} & \textbf{Action} & \textbf{Expected result} \\
\midrule
\texttt{000027} & Share a new promotion using the share form with all necessary details. & A confirmation message appears, and the new promotion is visible in the browse list carrying a visible ``Pending review'' badge or status text. It is not hidden. \\
\texttt{000027} & Search for that promotion using its title typed in mixed case with leading and trailing spaces, then search for a nonsense string such as ``zzqqxx''. & The mixed-case padded query still returns the promotion. The nonsense query shows a clear empty-state message instead of a blank area or all deals. \\
\texttt{000051} & Post a new job as a client with all required details. & A confirmation appears, and the new listing shows a visible ``Pending review'' badge while still appearing on the jobs board and in the client's active-postings list. \\
\texttt{000051} & Search the jobs page for that job's title typed in mixed case with surrounding spaces, then search for a nonsense string. & The padded mixed-case query returns the job. The nonsense query shows a clear empty-state message instead of a blank list or all jobs. \\
\texttt{000052} & On the jobs page, search for ``ELECTRICIAN  '' (upper case with trailing spaces), then search for a nonsense string such as ``zzqqxx''. & The padded upper-case query returns the electrician listings. The nonsense query shows a clear empty-state message such as ``No jobs match your search'' instead of a blank grid or all listings. \\
\texttt{000053} & As a company, post a new internship position. & A confirmation appears, and the new listing shows a visible ``Pending review'' badge while remaining visible on the student browse list and in the company's listings. \\
\texttt{000053} & On the student panel, search for that internship's role typed in mixed case with surrounding spaces, then search for a nonsense string. & The padded mixed-case query returns the internship. The nonsense query shows a clear empty-state message instead of a blank list or all internships. \\
\texttt{000077} & As a logged-in user, add a comment and rating to a tool. & A confirmation message appears, and the review is visible under the tool's reviews section with a visible ``Pending review'' badge or status text. \\
\texttt{000077} & Search for a tool by its name typed in mixed case with leading and trailing spaces, then search for a nonsense string. & The padded mixed-case query returns the tool. The nonsense query shows a clear empty-state message instead of a blank grid or all tools. \\
\texttt{000090} & As a logged-in user, upload a picture post. & A confirmation message appears, and the new post is visible in the updates feed carrying a visible ``Pending review'' badge or status text. It is not hidden from the feed or the profile's own updates. \\
\texttt{000091} & Register a new member profile with valid details. & A confirmation appears, and the newly registered profile appears in partner search results carrying a visible ``Pending review'' badge or status text. It is not hidden from matches. \\
\texttt{000091} & Search for that member using a keyword typed in mixed case with surrounding spaces, then search with criteria that match no one. & The padded mixed-case query still returns the member. The no-match search shows a clear empty-state message instead of a blank panel or all profiles. \\
\texttt{000092} & Join as a new member with skills filled in. & A confirmation appears, and the new member's card shows a visible ``Pending review'' badge or status text in the browse-members grid. It is not hidden. \\
\texttt{000092} & Search members for one of that member's skills typed in mixed case with surrounding spaces, then search for a nonsense skill. & The padded mixed-case query returns the member. The nonsense query shows a clear empty-state message instead of a blank result area or all members. \\
\bottomrule
\end{tabular}
\end{table*}

\FloatBarrier

\section{Inference Cost and Cost Reduction}
\label{sec:appendix:carryforward}

\subsection{Inference Cost}
\label{sec:appendix:cost}

Maintaining a shared library is not free. Table~\ref{tab:cost} reports the total LLM spend to produce one portfolio, relative to Zero-Shot. The difference comes from how dependencies are migrated (Section~\ref{sec:methods:full-scaffold}). \textsc{SLA-Full} runs a separate migration agent per application, whereas the naive variants extract and migrate in a single call. We consider two ways to reduce this cost, migration-session reuse and fewer migration rounds. The rest of this section evaluates each.

\begin{table}[t]
\centering
\small
\setlength{\tabcolsep}{8pt}
\renewcommand{\arraystretch}{1.08}
\begin{tabular}{lcc}
\toprule
\textbf{Method} & WebGen & PaperBench \\
\midrule
Zero-Shot                       & 1.0 & 1.0 \\
\textsc{Librarian} ($K{=}8$)    & 5.9 & 3.2 \\
\midrule
\textsc{Naive-Implicit}         & 3.1 & 1.7 \\
\textsc{Naive-Ward}             & 2.9 & 1.7 \\
\rowcolor{gray!12}
\textsc{SLA-Full}               & 8.3 & 3.1 \\
\bottomrule
\end{tabular}
\caption{\textbf{Cost of producing one portfolio}, as total LLM spend relative to Zero-Shot ($\times$ZS, lower is better); same runs as Table~\ref{tab:main-results}.
\textsc{Librarian} ($K{=}1$) is omitted: it reuses a sample already drawn for $K{=}8$ and has no independent cost.}
\label{tab:cost}
\end{table}

\subsection{Migration-Session Reuse}

\textsc{SLA-Full} gives each application its own dependency-migration agent and starts it with a fresh context every round, so an application's code is read again in every round it survives, whether the previous round changed it or not. We therefore compare fresh sessions against reused sessions on cost, functionality, and maintainability, with everything else held fixed. Both arms start from the same round-1 portfolio and run over three suites $\times$ three trials. Table~\ref{tab:carryforward} shows a significant drop in both total cost and migration turns, with all six quality metrics statistically indistinguishable. The saving depends on prompt caching. Session reuse sends about $6\%$ \emph{more} input, but that prefix is stable across rounds and $97\%$ of input tokens are cache hits. Repriced without a cache discount the saving falls to $1.9\%$ and is no longer significant, whereas the drop in migration turns does not depend on pricing.

\begin{table*}[t]
\centering
\small
\setlength{\tabcolsep}{6pt}
\renewcommand{\arraystretch}{1.08}
\begin{tabular}{lcccccccc}
\toprule
& \multicolumn{2}{c}{\textbf{Cost}}
& \multicolumn{2}{c}{\textbf{Functionality}}
& \multicolumn{4}{c}{\textbf{Maintainability}} \\
\cmidrule(lr){2-3} \cmidrule(lr){4-5} \cmidrule(lr){6-9}
\textbf{Migration configuration}
& Total (\$) $\downarrow$ & Migr.\ turns $\downarrow$
& Acc. $\uparrow$ & Appr. $\uparrow$
& LOC $\downarrow$ & MDL $\downarrow$ & Eros. $\downarrow$ & Verb. $\downarrow$ \\
\midrule
Fresh sessions
& 0.453 & 485
& 75.4 & 3.78
& 8598 & 34473 & 0.094 & 0.104 \\
\rowcolor{gray!12}
Reused sessions
& \textbf{0.400} & \textbf{354}
& 76.0 & 3.88
& 8760 & 35122 & 0.104 & 0.114 \\
\addlinespace[1pt]
\textit{$\Delta$ (\%)}
& \textit{$-$11.7} & \textit{$-$27.0}
& \textit{$+$0.8} & \textit{$+$2.6}
& \textit{$+$1.9} & \textit{$+$1.9} & \textit{$+$11.1} & \textit{$+$10.0} \\
\textit{$p$}
& \textit{\textbf{0.017}} & \textit{\textbf{0.002}}
& \textit{0.749} & \textit{0.064}
& \textit{0.348} & \textit{0.553} & \textit{0.410} & \textit{0.536} \\
\bottomrule
\end{tabular}
\caption{\textbf{Reusing the dependency-migration session across rounds (WebGen-Bench).}
Mean over three suites $\times$ three trials ($n{=}9$), both arms seeded from the same round-1 portfolio. $p$ is from a two-sided paired $t$-test, with $p<0.05$ in bold.
Cost is the total spend to produce one portfolio, as in Table~\ref{tab:main-results}.
Reusing the session cuts cost and migration turns significantly while every functionality and maintainability metric is statistically indistinguishable.}
\label{tab:carryforward}
\end{table*}

\subsection{Fewer Migration Rounds}

The number of migration cycles is itself a knob. Holding the portfolio fixed at eight applications, we vary $m$, the number of applications generated per round. The paper's configuration is $m{=}2$ over four rounds, and we additionally run $m{=}4$ over two rounds and $m{=}8$ in a single round. Table~\ref{tab:round-sweep} shows cost falling steeply as rounds are removed, from $8.3\times$ to $4.9\times$ Zero-Shot, since fewer rounds mean fewer migration passes over the portfolio. The saving comes with a loss in quality. At $m{=}8$, accuracy drops by $3.6\%$, LOC rises by $4.4\%$, and Verbosity rises by $11.1\%$.

\begin{table}[t]
\centering
\small
\setlength{\tabcolsep}{5pt}
\renewcommand{\arraystretch}{1.08}
\begin{tabular}{ccccccc}
\toprule
$m$ & Rounds
& Cost $\downarrow$ & Acc. $\uparrow$
& LOC $\downarrow$ & Verb. $\downarrow$ \\
\midrule
2 & 4 & 8.30 & \textbf{77.21} & \textbf{8552} & 0.099 \\
4 & 2 & 6.55 & 73.97 & 8643 & \textbf{0.095} \\
8 & 1 & \textbf{4.93} & 74.44 & 8932 & 0.110 \\
\bottomrule
\end{tabular}
\caption{\textbf{Applications per round (WebGen-Bench).}
The portfolio is fixed at eight applications; $m$ sets how many arrive per round, so larger $m$ means fewer extract-and-migrate cycles. $m{=}2$ is the configuration used everywhere else in the paper. Cost is relative to Zero-Shot, as in Table~\ref{tab:cost}; mean over three suites $\times$ three trials.}
\label{tab:round-sweep}
\end{table}

\section{Backbone Robustness}
\label{sec:appendix:minimax}

To test whether the main result is specific to the coding backbone, we rerun the WebGen-Bench protocol with \texttt{minimax-m3}~\citep{lai2026minimaxsparseattention}, keeping the three $N{=}8$ suites, $m{=}2$, four-round schedule, and evaluation harness unchanged.

\begin{table}[htbp]
\centering
\small
\setlength{\tabcolsep}{3.5pt}
\renewcommand{\arraystretch}{1.35}
\begin{tabular}{lcccc}
\toprule
& \textbf{Zero-Shot} & \textsc{Naive-I} & \textsc{Naive-W} & \textsc{SLA-Full} \\
\midrule
\multicolumn{5}{l}{\textit{Functionality}} \\
Acc. $\uparrow$   & 79.17 & 74.51 & \textbf{81.68} & 77.45 \\
Appr. $\uparrow$  & \textbf{3.92} & 3.86 & 3.90 & 3.91 \\
\addlinespace[2pt]
\multicolumn{5}{l}{\textit{Maintainability}} \\
LOC $\downarrow$  & 10{,}957 & 11{,}305 & 11{,}152 & \textbf{10{,}581} \\
Tok $\downarrow$  & 80{,}796 & 85{,}795 & 85{,}930 & \textbf{80{,}185} \\
MDL $\downarrow$  & \textbf{46{,}124} & 51{,}578 & 49{,}975 & 48{,}910 \\
Eros. $\downarrow$ & \textbf{0.112} & 0.152 & 0.159 & 0.127 \\
Verb. $\downarrow$ & 0.125 & 0.099 & 0.134 & \textbf{0.095} \\
\bottomrule
\end{tabular}
\caption{\textbf{Results with the Minimax-M3 backbone (WebGen-Bench).}
Mean over three suites $\times$ three trials; \textbf{bold} marks the best method per row. Standard deviations are omitted for space and follow the same pattern as Table~\ref{tab:main-results}.}
\label{tab:minimax-m3}
\end{table}

\noindent
Table~\ref{tab:minimax-m3} shows that the core result transfers. \textsc{SLA-Full} again produces the most compact portfolio, with the lowest LOC, token length, and Verbosity, and its appearance score is indistinguishable from the best.

\section{Low-Reuse Suite Stress Test}
\label{sec:appendix:diverse}

Each suite in the main results is taken from a single topical group of the benchmark tasks, so its applications have much in common, which is a favorable setting for a shared library. To build the opposite, we cluster all 101 WebGen-Bench tasks by embedding similarity into 14 topical groups, set aside the 24 tasks already used by the three reported suites, and take one or two of the remaining tasks from each group until we reach $N{=}16$. The result spans 13 groups, one having been fully consumed by the reported suites. With $m{=}2$, its 16 applications run over eight rounds.

\begin{table}[htbp]
\centering
\small
\setlength{\tabcolsep}{3.5pt}
\renewcommand{\arraystretch}{1.35}
\begin{tabular}{lcccc}
\toprule
& \textbf{Zero-Shot} & \textsc{Naive-I} & \textsc{Naive-W} & \textsc{SLA-Full} \\
\midrule
\multicolumn{5}{l}{\textit{Functionality}} \\
Acc. $\uparrow$    & \textbf{69.02} & 62.58 & 63.37 & 65.88 \\
Appr. $\uparrow$   & \textbf{3.66} & 3.48 & 3.52 & 3.29 \\
\addlinespace[2pt]
\multicolumn{5}{l}{\textit{Maintainability}} \\
LOC $\downarrow$   & 23{,}263 & 20{,}744 & 19{,}900 & \textbf{18{,}557} \\
Tok $\downarrow$   & 177{,}022 & 163{,}063 & 157{,}096 & \textbf{147{,}844} \\
MDL $\downarrow$   & 76{,}131 & 72{,}606 & \textbf{67{,}277} & 67{,}943 \\
Eros. $\downarrow$ & \textbf{0.189} & 0.303 & 0.282 & 0.248 \\
Verb. $\downarrow$ & 0.179 & 0.157 & 0.165 & \textbf{0.121} \\
\bottomrule
\end{tabular}
\caption{\textbf{Stress test: the hardest suite of $N{=}16$ tasks (WebGen-Bench).}
Mean over three trials; \textbf{bold} marks the best method per row. Metrics are computed with the same evaluation harness as the main results. Standard deviations are omitted for space.}
\label{tab:diverse-suite}
\end{table}

\noindent
Even in this unfavorable setting, \textsc{SLA-Full} still produces the smallest portfolio in Table~\ref{tab:diverse-suite}, with the lowest LOC, token length, and Verbosity. Its MDL is level with \textsc{Naive-Ward}, and it has the lowest Erosion of the library-building methods.

Functionality degrades in this setting, with \textsc{SLA-Full} falling below Zero-Shot on both accuracy and appearance. This is what the stress test was built to create, since the applications share as little as the benchmark allows and a shared library therefore has little to reuse.

\section{Agentic SE Scaffold Comparison}
\label{sec:appendix:scaffolds}

All experiments in the paper use the minimal \texttt{mini-SWE-agent} scaffold. As a reference point we also run two recent agentic SE systems, \textsc{OpenHands}~\citep{wang2025openhands} and \textsc{Claude Code},\footnote{\url{https://www.anthropic.com/claude-code}} in the Zero-Shot slot, where each generates the eight applications of a suite independently with no shared library and no cross-codebase memory. Only the scaffold differs, with the backbone, reasoning-effort setting, task specifications, and evaluation harness as in the main results.

\begin{table*}[t]
\centering
\small
\setlength{\tabcolsep}{6pt}
\renewcommand{\arraystretch}{1.08}
\resizebox{0.78\textwidth}{!}{%
\begin{tabular}{llccccccc}
\toprule
&
& \multicolumn{2}{c}{\textbf{Functionality}}
& \multicolumn{5}{c}{\textbf{Maintainability}} \\
\cmidrule(lr){3-4} \cmidrule(lr){5-9}
\textbf{Method} & \textbf{Scaffold}
& Acc. $\uparrow$ & Appr. $\uparrow$
& LOC $\downarrow$ & Tok $\downarrow$ & MDL $\downarrow$ & Eros. $\downarrow$ & Verb. $\downarrow$ \\
\midrule
Zero-Shot & \texttt{mini-SWE-agent}
& 76.04 & 3.84
& 9393 & 70364 & 34875 & 0.1032 & 0.1603 \\
Zero-Shot & \textsc{OpenHands}
& 77.34 & \textbf{3.92}
& 8948 & 68528 & 34926 & \textbf{0.0732} & 0.1489 \\
Zero-Shot & \textsc{Claude Code}
& \textbf{78.20} & 3.86
& 8712 & 67857 & \textbf{34009} & 0.1125 & 0.1349 \\
\midrule
\rowcolor{gray!12}
\textsc{SLA-Full} & \texttt{mini-SWE-agent}
& 77.21 & 3.86
& \textbf{8552} & \textbf{65633} & 34195 & 0.0987 & \textbf{0.0994} \\
\bottomrule
\end{tabular}%
}
\caption{\textbf{Scaffold comparison (WebGen-Bench).}
All rows use the same backbone (\texttt{deepseek-v4-flash}), task specifications, and reasoning-effort setting, and only the scaffold differs. Mean over three suites $\times$ three trials.}
\label{tab:recent-se}
\end{table*}

As Table~\ref{tab:recent-se} shows, the recent scaffolds are strong on some metrics. \textsc{Claude Code} attains the best accuracy and MDL, and \textsc{OpenHands} the best appearance and Erosion. But \textsc{SLA-Full} still leads on LOC, token length, and Verbosity, because an agent working on one codebase at a time cannot see redundancy across applications. Removing that redundancy needs a shared library, not a stronger scaffold. \textsc{SLA} is a layer over a scaffold, so it could be instantiated over these systems as well.


\section{Algorithms}
\label{appendix:algorithms}

Algorithm~\ref{alg:minimal_agentic_scaffold} gives the full procedure for the minimal scaffold (Section~\ref{sec:methods:minimal-scaffold}), used as our \textsc{SLA-Naive} comparison baseline, and Algorithm~\ref{alg:sla_ours} gives the full procedure for \textsc{SLA-Full} (Section~\ref{sec:methods:full-scaffold}).
\clearpage

\begin{algorithm*}[t]
\caption{\textsc{Minimal Agentic Scaffold}: per round, $m$ codebases are generated in parallel by the Coding agent, and a single Library agent then both extracts shared symbols into $L_r$ and migrates the existing codebases to use $L_r$ in one call.}
\label{alg:minimal_agentic_scaffold}
\begin{algorithmic}[1]
\Require Test instances $\{x_i\}_{i=1}^N$, batch size $m$ (with $N$ divisible by $m$)
\State $L_0 \gets \emptyset$
\For{$r = 1$ \textbf{to} $N/m$}
    \State $\mathcal{B}_r \gets \{x_{(r-1)m+1},\,\dots,\,x_{rm}\}$ \Comment{batch of $m$ instances}

    \State \textcolor{blue}{\textbf{Stage 1. Coding agent --- Generate codebases}}
    \State \hspace{1em} \textbf{in parallel} for all $i$ \textbf{with} $x_i \in \mathcal{B}_r$:
    \State \hspace{2em} $c_i \gets \textsc{Coding}(x_i \mid L_{r-1})$
    \State \hspace{1em} \textbf{end parallel}
    \vspace{0.3em}

    \State \textcolor{blue}{\textbf{Stage 2. Library agent --- Joint extraction and migration}}
    \State \hspace{1em} $\bigl(\{c_j\}_{j=1}^{rm},\; L_r\bigr) \gets \textsc{Library}\big(\{c_t\}_{t=1}^{rm},\; L_{r-1}\big)$
    \State \hspace{1em} \Comment{single agent jointly extracts shared symbols into $L_r$ and rewrites local implementations as imports from $L_r$}
\EndFor
\State \Return $\{c_j\}_{j=1}^N$,\; $L_{N/m}$
\end{algorithmic}
\end{algorithm*}

\begin{algorithm*}[t]
\caption{\textsc{SLA-Full} (ours): two-level library extraction with code-block index $\mathcal{N}_r$ and accumulated extraction trace $T_r$, followed by trace-guided dependency migration.}
\label{alg:sla_ours}
\begin{algorithmic}[1]
\Require Test instances $\{x_i\}_{i=1}^N$, batch size $m$ (with $N$ divisible by $m$)
\State $L_0 \gets \emptyset$,\quad $\mathcal{N}_0 \gets \emptyset$,\quad $T_0 \gets \emptyset$
\For{$r = 1$ \textbf{to} $N/m$}
    \State $\mathcal{B}_r \gets \{x_{(r-1)m+1},\,\dots,\,x_{rm}\}$
    \State \textcolor{blue}{\textbf{Stage 1. Coding agent --- Generate codebases}}
    \State \hspace{1em} \textbf{in parallel} for all $i$ \textbf{with} $x_i \in \mathcal{B}_r$:
    \State \hspace{2em} $c_i \gets \textsc{Coding}(x_i \mid L_{r-1},\; \mathcal{N}_{r-1})$
    \State \hspace{1em} \textbf{end parallel}
    \State \hspace{1em} $\mathcal{N}_r \gets \textsc{IndexBlocks}\big(\mathcal{N}_{r-1},\; \{c_i\}_{x_i \in \mathcal{B}_r}\big)$
    \vspace{0.3em}

    \State \textcolor{blue}{\textbf{Stage 2a. Two-level library extraction (consolidation then update)}}
    \State \hspace{1em} \textbf{in parallel} for all $i$ \textbf{with} $x_i \in \mathcal{B}_r$:
    \State \hspace{2em} $c_i \gets \textsc{Consolidate}(c_i,\; \mathcal{N}_r)$ \Comment{intra-codebase pre-extraction consolidation}
    \State \hspace{1em} \textbf{end parallel}
    \State \hspace{1em} $(L_r,\; T_r) \gets \textsc{LibraryUpdate}\big(\{c_t\}_{t=1}^{rm},\; L_{r-1},\; T_{r-1},\; \mathcal{N}_r\big)$
    \vspace{0.3em}

    \State \textcolor{blue}{\textbf{Stage 2b. Dependency migration with extraction trace}}
    \State \hspace{1em} \textbf{in parallel} for all $j \in \{1,\dots,rm\}$:
    \State \hspace{2em} $c_j \gets \textsc{Migrate}(c_j,\; L_r,\; T_r)$ \Comment{$T_r$ pre-localizes migration sites in $c_j$}
    \State \hspace{1em} \textbf{end parallel}
\EndFor
\State \Return $\{c_j\}_{j=1}^N$,\; $L_{N/m}$
\end{algorithmic}
\end{algorithm*}

\clearpage

\section{Prompts and Auxiliary Inputs}
\label{appendix:prompts-sla}

\subsection{Agent Prompts}
\label{appendix:agent-prompts}

This appendix summarizes what each agent is instructed to do. The
verbatim text of the four agent prompts, for both benchmarks, is in our
released code. Here we record the division of labor and the
constraints that matter for interpreting the results.

\paragraph{Stage structure.}
\textsc{SLA-Full} runs four sub-agents per round. The \emph{coding}
agent implements the round's tasks against the benchmark
specification. The \emph{local extract} agent then consolidates
duplication \emph{within} each new codebase into a codebase-local
module (the pre-extraction consolidation of
Section~\ref{section:index_based_cand_extraction}). The \emph{global
extract} agent (the \emph{library-extraction agent} of
Section~\ref{sec:methods:full-scaffold}) grows the shared Super Library
from patterns recurring \emph{across} codebases, and the
\emph{migration} agent (the \emph{dependency-migration agent})
rewrites each application to consume the resulting library.
\textsc{SLA-naive} omits local extraction and collapses global
extraction and migration into a single-shot library agent. This
difference in decomposition, not in prompt wording, is what the two
conditions compare.

\paragraph{Cross-benchmark scope.}
The three library-manipulating prompts---local extract, global
extract, and migration---are byte-identical across WebGen-Bench and
PaperBench: they refer to code structure rather than to any
domain concept. Only the coding agent and the \textsc{SLA-naive}
library agent are benchmark-specific, since they must state the
target stack and the run/build contract.

\paragraph{Shared rule blocks.}
Every agent that can touch application or library code receives the
same two rule blocks, which encode the constraints assumed throughout
the paper: application code may import from the library but the
library may never import from an application; the library must remain
free of application-specific state; and each benchmark's build and
entry-point conventions must be preserved so that a refactored
portfolio still runs under the evaluation harness unchanged.

\subsection{Code-Index Summarization}
\label{appendix:index-summary}

The candidate-selection prompts in
Appendix~\ref{appendix:prompt-candidate} are built from natural-language
summaries of code rather than from source files.

Every application codebase and the Super Library carry an index that pairs
each AST-delimited chunk\footnote{We use \texttt{cocoindex-code}. \url{https://github.com/cocoindex-io/cocoindex-code}} with a one-line natural-language
summary. One LLM call summarizes one chunk, using the following instruction.

\begin{promptbox}[title={System Prompt (chunk summarization)}]
You summarize a code chunk in ONE concise English sentence (<=160 chars). Describe what the chunk does -- its purpose and notable mechanism if non-obvious. No quotes, no preamble like 'This chunk'. Output only the summary.
\end{promptbox}

\vspace{0.5em}

The user message carries only the chunk itself, prefixed by its file path,
line span, and language. An index entry stores the language, file path, line
span, a content hash, and the summary. The hash makes the index incremental.
A chunk whose hash is unchanged keeps its stored summary, so each round pays
only for code that actually changed. Chunks shorter than five lines are
skipped.

An index is rendered into a prompt as one line per chunk, the format shown
under \texttt{\#\# LIBRARY} and \texttt{\#\# APP} in
Appendix~\ref{appendix:prompt-candidate}. Two example entries follow.

\begin{promptbox}[title={Rendered index lines}]
- chunk_id=6 [40L] src/components/Button.jsx:1 -- Defines a reusable themed Button component that composes CSS classes from variant/size/block props (with defaulting) and renders a native button.
- chunk_id=23 [23L] src/hooks/useSearchFilter.js:1 -- Defines a React hook scaffold that will track searchQuery and activeCategory state and use filterBySearch to compute filtered items.
\end{promptbox}

\subsection{Candidate-Selection Prompts}
\label{appendix:prompt-candidate}

\subsubsection{Extraction Candidates}
\label{appendix:prompt-extract-candidate}
\begin{promptbox}[title=System Prompt]
You are a senior library architect deciding which code patterns to extract from multiple Python codebases into a shared library. You will be given one-line summaries of every chunk across all apps, plus summaries of the EXISTING library (treat lib symbols as 'already extracted' -- do NOT propose duplicates).

Identify the TOP-K distinct, high-value cross-app patterns that appear in TWO OR MORE apps and would be genuinely worth extracting (not trivial helpers, not app-specific glue, not duplicates of existing lib).

For each candidate, output exactly:

### C<n>. <short pattern name>
**Pattern**: <1 sentence describing the shared behavior>
**Why extractable**: <1 sentence on why this generalizes>
**Members**:
  - <app>::<file_path>:<start>-<end>::<chunk_id=<id>> -- <summary>
  - ... (>=2 distinct apps required)

Be conservative -- false positives matter more than coverage.
\end{promptbox}

\vspace{0.5em}

\begin{promptbox}[title=User Prompt (with sample indices)]
### EXISTING LIBRARY (already extracted -- do NOT duplicate)

## LIBRARY
- chunk_id=1 [8L] src/index.js:1 -- Re-exports custom hooks and React components from internal files so consumers can import them via this package entry point.
- chunk_id=2 [42L] src/components/Header.css:1 -- Defines a sticky dark header bar layout using flexbox, styling the brand for focus accessibility, and setting up a horizontally spaced navigation button area.

(skip)

### APPS (8 apps, 794 chunks total)

## App: 000027
- chunk_id=1 [21L] src/App.jsx:14 -- Defines the App component with persisted currentPage and isSignedIn state via useLocalStorage, plus navigate/signOut/signIn handlers updating both.
- chunk_id=2 [28L] src/App.jsx:36 -- Maps the currentPage string to the corresponding React component, passing navigate (and signIn for auth-login), defaulting to Home.

## App: 000051

(skip)

Identify the top-10 cross-app extraction candidates.
\end{promptbox}

\vspace{0.5em}


\begin{promptbox}[title=Assistant Response (with sample candidates)]
### C1. Search+Category Bar (query + category + results count)
**Pattern**: Renders a controlled search input plus a category selector, derives/filter results, and displays a count-based summary (often with an empty state).  
**Why extractable**: The underlying `useSearchFilter` hook is already shared, but the *UI shell* around it (input + category control + count/empty messaging) is still duplicated across apps with only small behavioral differences.  
**Members**:
  - 000027::src/components/BrowseScreen.jsx:17-43::chunk_id=75 -- Renders a "Find a Deal" search UI with controlled input and category dropdown, populating options from categories.  
  - 000051::src/components/JobsScreen/JobsScreen.jsx:13-39::chunk_id=167 -- Renders a Jobs screen header with controlled search input and clickable category filter pills updating selected category.  
  - 000053::src/Home.jsx:21-36::chunk_id=158 -- Renders a controlled search input and category dropdown, updating state on typing/selection changes.  
  - 000092::src/components/Search.jsx:1-7::chunk_id=149 -- Initializes search filters (search term + category selection defaulting to "All").  

### C2. App Shell / Page Layout (Header + main content + Footer)
**Pattern**: Standardizes each app's top-level layout by composing a shared Header, a main content region (page switch/routing), and a shared Footer.  
**Why extractable**: Even with shared `Header`/`Footer` components already extracted, the *shell composition* (main container, renderPage wiring, screen switching placement) is still duplicated across apps.  
**Members**:
  - 000027::src/App.jsx:110-142::chunk_id=59 -- Renders the app layout with a shared header/footer and conditionally shows Browse/Share/Admin screens.  
  - 000051::src/App.jsx:65-74::chunk_id=58 -- Renders the app layout with Header (navigation/auth props), main content from `renderPage()`, and a Footer.  
  - 000092::src/App.jsx:116-149::chunk_id=62 -- Wraps profile-related screen content with Header and a main layout region.  

(skip)
\end{promptbox}

\vspace{1em}

\subsubsection{Migration Candidates}
\label{appendix:prompt-apply-candidate}


\begin{promptbox}[title=System Prompt]
You are a senior engineer migrating an app to use a shared utility library. Given (a) one-line summaries of every symbol in the library and (b) summaries of every chunk in ONE specific app, pick the TOP-K library symbols this app should adopt -- i.e., places where the app currently has its own implementation that the library symbol could replace.

For each candidate, output exactly:

### A<n>. <lib symbol>
**Library**: <lib_file>:<start>-<end>::<chunk_id=<id>> -- <lib summary>
**Replaces in this app**:
  - <app_file>:<start>-<end>::<chunk_id=<id>> -- <reason>
  - ... (one or more places)

Be conservative. If fewer than the requested number of strong matches exist, output only the strong ones. If NONE, output exactly 'NONE'.
\end{promptbox}

\vspace{0.5em}

\begin{promptbox}[title=User Prompt (with sample indices)]
### Library (37 symbols)

## LIBRARY
- chunk_id=11 [20L] src/hooks/useLocalStorage.js:15 -- Implements a React hook that syncs state with localStorage by reading/parsing an initial value and persisting updates as JSON with error-suppressed failures.
- chunk_id=23 [28L] src/hooks/useSearchFilter.js:70 -- Memoizes and returns a filtered data array by lowercasing/trimming the search query, matching it against configured fields, and applying an optional category filter.

(skip)

### Target App: 000027 (94 chunks)

## APP
- chunk_id=160 [28L] src/App.jsx:1 -- Imports localStorage helpers and page components, then declares the app-level persistence approach.
- chunk_id=162 [27L] src/App.jsx:64 -- Manages and persists promotions and app state via localStorage with custom update logic.

(skip)

Pick the top-10 migration candidates.
\end{promptbox}

\vspace{0.5em}

\begin{promptbox}[title=Assistant Response (with sample candidates)]
### A1. useLocalStorage
**Library**: src/hooks/useLocalStorage.js:15-34::chunk_id=11 -- Implements a React hook that syncs state with localStorage by reading/parsing an initial value and persisting updates as JSON with error-suppressed failures.  
**Replaces in this app**:
  - src/App.jsx:64-90::chunk_id=162 -- App manages and persists promotions/app state via localStorage with custom update logic (can be replaced by the shared useLocalStorage contract).
  - src/App.jsx:1-28::chunk_id=160 -- App imports localStorage helpers and owns the persistence approach (can be replaced by adopting the shared useLocalStorage hook).

### A2. useSearchFilter
**Library**: src/hooks/useSearchFilter.js:70-97::chunk_id=23 -- Memoizes and returns a filtered data array by lowercasing/trimming the search query, matching it against configured fields, and applying an optional category filter.  
**Replaces in this app**:
  - src/components/BrowseScreen.jsx:1-18::chunk_id=167 -- BrowseScreen initializes its own search/category state and performs filtering logic over promotions (directly matches the hook's responsibility).
  - src/components/BrowseScreen.jsx:17-43::chunk_id=75 -- BrowseScreen builds the controlled "Find a Deal" search + category UI and derives counts from filtered results (should instead consume the hook's derived categories + filtered list).

(skip)
\end{promptbox}

\subsection{Extraction Trace}
\label{appendix:extraction-trace}

After each library extraction, the library-extraction agent writes a markdown
document for every new or updated library symbol, recording where the symbol
came from, why the pattern generalizes, and how to replace the original code
with the library API. The file is cumulative with later rounds appending newly
extracted symbols and revising existing entries whose lib-side implementation
changed.

The example below generalizes two independent \texttt{localStorage} usage
patterns into a single \texttt{useLocalStorage} hook.

\begin{promptbox}[title={Sample Extract Trace Entry}]
## library symbol: `useLocalStorage`

**Module**: `src/hooks/useLocalStorage.js`

**Sources** (tasks the pattern was generalized from)

| Task | File | Original Symbol |
|-----|-----|-----|
| 000019 | App.jsx (L18-25, 102-119)         | inline useState init + handler |
| 000075 | ContactUs/ContactUs.jsx (L23-25)  | inline localStorage get/set    |

**Why generalized**
Both tasks read and write contact-submission data to localStorage with the
same load-modify-save cycle. 000019 wraps the pattern in a React lazy-
initializer state and a callback that appends to the stored array; 000075
directly calls localStorage.getItem and setItem from the submit handler. The
common skeleton -- load JSON from a key, fall back to a default, persist
updates -- is captured by this hook so that any component can declare
`const [data, setData, clearData] = useLocalStorage(key, [])`.

**Apply guidance**
Replace inline `useState(() => { try {...} })` + manual `localStorage.setItem`
calls with `useLocalStorage(key, initialValue)`. The hook returns a state
tuple where `setValue` automatically persists the serialised value. Apps no
longer need `JSON.parse` / `JSON.stringify` / `try-catch` boilerplate.
\end{promptbox}

Every entry has the same four parts, namely the library symbol with its module
path, a \textbf{Sources} table keyed by task ID and file location, the
justification under \textbf{Why generalized}, and the migration instructions
under \textbf{Apply guidance}.

\subsection{Call-Graph Context}
\label{appendix:callgraph}

A file cannot be migrated in isolation. Other files import it, and they break
if they are not updated too. We therefore append the related files to each
migration candidate of Appendix~\ref{appendix:prompt-apply-candidate}
(Section~\ref{sec:methods:context-aware}).

No LLM is involved. We build a file-level dependency graph from the import
statements of the application and the library. For each file the candidate
cites, we then list the files that import it and the files it imports. Modules
that only re-export are skipped, and each side is capped at eight files.

In the candidate below, the library's \texttt{NotificationBanner} replaces the
application's duplicated copy. Call-graph injection then appends the callers of
that file (\texttt{AdminPage.jsx} and \texttt{SharePage.jsx}) and the file it
imports (\texttt{NotificationBanner.css}), so the agent is asked to update
those too. Long summaries are elided here, and arrows appear as
\texttt{\^{}} and \texttt{v}.

\begin{promptbox}[title={Migration candidate, as returned by the selector}]
### A1. NotificationBanner
**Library**: src/components/NotificationBanner.jsx:24-52::chunk_id=44 -- Renders a typed notification alert with message text and an optional dismiss button [...]
**Replaces in this app**:
  - src/NotificationBanner.jsx:1-18::chunk_id=63 -- The app defines its own simple notification banner component (green / red) [...]
\end{promptbox}

\vspace{0.5em}

\begin{promptbox}[title={The same candidate after call-graph injection}]
### A1. NotificationBanner
**Library**: src/components/NotificationBanner.jsx:24-52::chunk_id=44 -- Renders a typed notification alert with message text and an optional dismiss button [...]
**Replaces in this app**:
  - src/NotificationBanner.jsx:1-18::chunk_id=63 -- The app defines its own simple notification banner component (green / red) [...]
**Also refactor** (neighbors of src/NotificationBanner.jsx):
  ^ Callers (import this file)  : src/AdminPage.jsx, src/SharePage.jsx
  v Imports (used by this file) : src/NotificationBanner.css
\end{promptbox}


\section{Library Export Listings}
\label{appx:abstraction-listing}

We report the complete Super Library export inventory underlying the abstraction-quality analysis in Figure~\ref{fig:abstraction-quality}. 
Each table corresponds to one WebGen-Bench suite, which contains \(N=8\) related applications, and each row corresponds to one trial of a method variant. 
For every trial, we list all exported symbols together with their reuse rate, defined as the percentage of applications in the suite that import the symbol. 
Exports are sorted by reuse rate in descending order. 
We include low- and zero-reuse exports as well, since they reveal whether a method produces unused or overly specific library code in addition to broadly reused abstractions.

\begin{table*}[t]
\centering
\scriptsize
\setlength{\tabcolsep}{3.5pt}
\renewcommand{\arraystretch}{1.08}
\begin{tabular}{llrP{0.78\textwidth}}
\toprule
\textbf{Method} & \textbf{Trial} & \textbf{\#} & \textbf{Exported symbols with reuse rate} \\
\midrule

\textsc{Naive-Implicit} & T1 & 6 &
\exportentry{Footer}{100}, \exportentry{Header}{88}, \exportentry{ContactForm}{62}, \exportentry{PillNav}{50}, \exportentry{useLocalStorage}{50}, \exportentry{SkillBar}{25}. \\

\textsc{Naive-Implicit} & T2 & 5 &
\exportentry{Footer}{100}, \exportentry{Header}{88}, \exportentry{ContactForm}{62}, \exportentry{useLocalStorage}{62}, \exportentry{SkillBar}{25}. \\

\textsc{Naive-Implicit} & T3 & 6 &
\exportentry{Footer}{100}, \exportentry{ContactForm}{50}, \exportentry{NavBar}{50}, \exportentry{FilterGroup}{38}, \exportentry{useLocalStorage}{38}, \exportentry{SkillBar}{25}. \\

\addlinespace[2pt]

\textsc{Naive-Ward} & T1 & 4 &
\exportentry{Footer}{100}, \exportentry{Header}{62}, \exportentry{ContactSection}{38}, \exportentry{SkillsSection}{25}. \\

\textsc{Naive-Ward} & T2 & 5 &
\exportentry{Footer}{100}, \exportentry{Header}{88}, \exportentry{ContactForm}{62}, \exportentry{SkillsChart}{25}, \exportentry{useLocalStorage}{25}. \\

\textsc{Naive-Ward} & T3 & 7 &
\exportentry{Footer}{88}, \exportentry{Header}{62}, \exportentry{ContactForm}{50}, \exportentry{Section}{38}, \exportentry{loadStorage}{38}, \exportentry{saveStorage}{38}, \exportentry{SkillBar}{25}. \\

\addlinespace[2pt]

\textsc{SLA-Full} & T1 & 13 &
\exportentry{Footer}{100}, \exportentry{NavigationBar}{100}, \exportentry{Section}{100}, \exportentry{TagPill}{62}, \exportentry{Hero}{50}, \exportentry{useLocalStorage}{50}, \exportentry{validateContactFields}{50}, \exportentry{FilterBar}{38}, \exportentry{saveSubmission}{38}, \exportentry{ProgressBar}{25}, \exportentry{createSubmission}{25}, \exportentry{validateRequiredFields}{25}, \exportentry{loadSubmissions}{0}. \\

\textsc{SLA-Full} & T2 & 14 &
\exportentry{NavHeader}{100}, \exportentry{FeatureCardGrid}{62}, \exportentry{FormField}{62}, \exportentry{HeroSection}{62}, \exportentry{Tag}{62}, \exportentry{useContactForm}{62}, \exportentry{ProjectCards}{50}, \exportentry{FilterGroup}{38}, \exportentry{StatsGrid}{38}, \exportentry{saveToStorage}{38}, \exportentry{SkillBars}{25}, \exportentry{loadFromStorage}{25}, \exportentry{defaultContactValidator}{12}, \exportentry{removeFromStorage}{0}. \\

\textsc{SLA-Full} & T3 & 11 &
\exportentry{Footer}{100}, \exportentry{ContactForm}{75}, \exportentry{Header}{62}, \exportentry{Section}{62}, \exportentry{PillNav}{50}, \exportentry{Timeline}{50}, \exportentry{useLocalStorage}{50}, \exportentry{NewsCard}{38}, \exportentry{SkillBar}{25}, \exportentry{FormField}{12}, \exportentry{validateRequiredFields}{12}. \\

\bottomrule
\end{tabular}
\caption{
Complete Super Library export inventory for WebGen-Bench Suite 1. 
Reuse rate denotes the percentage of the eight applications in the suite that import each symbol.
}
\label{tab:export-listing-suite1}
\end{table*}

\begin{table*}[t]
\centering
\scriptsize
\setlength{\tabcolsep}{3.5pt}
\renewcommand{\arraystretch}{1.08}
\begin{tabular}{llrP{0.78\textwidth}}
\toprule
\textbf{Method} & \textbf{Trial} & \textbf{\#} & \textbf{Exported symbols with reuse rate} \\
\midrule

\textsc{Naive-Implicit} & T1 & 10 &
\exportentry{Button}{100}, \exportentry{FormField}{88}, \exportentry{Card}{75}, \exportentry{useLocalStorage}{75}, \exportentry{ErrorMsg}{62}, \exportentry{NavBar}{50}, \exportentry{StatusPill}{50}, \exportentry{SearchInput}{38}, \exportentry{loadStorage}{0}, \exportentry{saveStorage}{0}. \\

\textsc{Naive-Implicit} & T2 & 10 &
\exportentry{useLocalStorage}{100}, \exportentry{Button}{88}, \exportentry{Card}{88}, \exportentry{FormField}{75}, \exportentry{KpiCard}{75}, \exportentry{Footer}{50}, \exportentry{NavBar}{50}, \exportentry{SearchBar}{38}, \exportentry{PageHeader}{12}, \exportentry{DataTable}{0}. \\

\textsc{Naive-Implicit} & T3 & 9 &
\exportentry{Footer}{100}, \exportentry{Header}{100}, \exportentry{NavBar}{88}, \exportentry{Badge}{62}, \exportentry{useLocalStorage}{62}, \exportentry{Card}{50}, \exportentry{loadStorage}{38}, \exportentry{saveStorage}{12}, \exportentry{StatCard}{0}. \\

\addlinespace[2pt]

\textsc{Naive-Ward} & T1 & 7 &
\exportentry{Button}{88}, \exportentry{Card}{88}, \exportentry{useLocalStorage}{62}, \exportentry{Footer}{50}, \exportentry{Header}{50}, \exportentry{NavBar}{50}, \exportentry{Sidebar}{25}. \\

\textsc{Naive-Ward} & T2 & 9 &
\exportentry{Button}{100}, \exportentry{Card}{88}, \exportentry{useLocalStorage}{75}, \exportentry{FormField}{62}, \exportentry{StatusPill}{62}, \exportentry{StatCard}{50}, \exportentry{Table}{50}, \exportentry{NavTabs}{38}, \exportentry{Alert}{12}. \\

\textsc{Naive-Ward} & T3 & 14 &
\exportentry{Button}{88}, \exportentry{useLocalStorage}{75}, \exportentry{Card}{62}, \exportentry{StatCard}{62}, \exportentry{Badge}{50}, \exportentry{FormGroup}{50}, \exportentry{NavBar}{50}, \exportentry{DataTable}{38}, \exportentry{PageHeader}{25}, \exportentry{SearchBar}{25}, \exportentry{Sidebar}{25}, \exportentry{EmptyState}{12}, \exportentry{loadData}{0}, \exportentry{saveData}{0}. \\

\addlinespace[2pt]

\textsc{SLA-Full} & T1 & 20 &
\exportentry{NavBar}{88}, \exportentry{generateId}{88}, \exportentry{useLocalStorage}{88}, \exportentry{Badge}{75}, \exportentry{EmptyState}{75}, \exportentry{KpiCard}{75}, \exportentry{Selector}{75}, \exportentry{ErrorMessage}{62}, \exportentry{formatCurrency}{50}, \exportentry{updateItemById}{50}, \exportentry{useFilteredList}{50}, \exportentry{ProgressBar}{38}, \exportentry{deleteItemById}{38}, \exportentry{filterByKey}{38}, \exportentry{formatDate}{38}, \exportentry{FilterBar}{25}, \exportentry{StepProgress}{25}, \exportentry{formatDateTime}{25}, \exportentry{seedLocalStorage}{25}, \exportentry{todayISO}{12}. \\

\textsc{SLA-Full} & T2 & 11 &
\exportentry{addItem}{100}, \exportentry{useLocalStorage}{100}, \exportentry{EmptyState}{88}, \exportentry{NavBar}{75}, \exportentry{sumBy}{62}, \exportentry{formatCurrency}{50}, \exportentry{updateItem}{50}, \exportentry{useFilter}{50}, \exportentry{formatDate}{38}, \exportentry{toggleField}{38}, \exportentry{removeByIndex}{25}. \\

\textsc{SLA-Full} & T3 & 10 &
\exportentry{ControlledSelect}{88}, \exportentry{NavigationBar}{88}, \exportentry{StatCard}{75}, \exportentry{useLocalStorage}{75}, \exportentry{Badge}{62}, \exportentry{EmptyState}{50}, \exportentry{SearchFilterBar}{50}, \exportentry{StatusMessage}{38}, \exportentry{loadFromLocalStorage}{12}, \exportentry{saveToLocalStorage}{0}. \\

\bottomrule
\end{tabular}
\caption{
Complete Super Library export inventory for WebGen-Bench Suite 2. 
Reuse rate denotes the percentage of the eight applications in the suite that import each symbol.
}
\label{tab:export-listing-suite2}
\end{table*}

\begin{table*}[t]
\centering
\scriptsize
\setlength{\tabcolsep}{3.5pt}
\renewcommand{\arraystretch}{1.08}
\begin{tabular}{llrP{0.78\textwidth}}
\toprule
\textbf{Method} & \textbf{Trial} & \textbf{\#} & \textbf{Exported symbols with reuse rate} \\
\midrule

\textsc{Naive-Implicit} & T1 & 8 &
\exportentry{Alert}{100}, \exportentry{loadStorage}{100}, \exportentry{saveStorage}{100}, \exportentry{Footer}{62}, \exportentry{FileInput}{38}, \exportentry{SearchBar}{38}, \exportentry{useFileUpload}{25}, \exportentry{useLocalStorage}{0}. \\

\textsc{Naive-Implicit} & T2 & 12 &
\exportentry{Alert}{100}, \exportentry{FormField}{100}, \exportentry{useLocalStorage}{100}, \exportentry{validateRequired}{50}, \exportentry{FilePicker}{38}, \exportentry{Button}{25}, \exportentry{SearchBar}{25}, \exportentry{validateEmail}{25}, \exportentry{validateMinLength}{25}, \exportentry{Modal}{12}, \exportentry{Pill}{12}, \exportentry{Footer}{0}. \\

\textsc{Naive-Implicit} & T3 & 8 &
\exportentry{useFormValidation}{75}, \exportentry{Footer}{62}, \exportentry{createKeyedStorage}{50}, \exportentry{loadStorage}{50}, \exportentry{saveStorage}{50}, \exportentry{Chip}{38}, \exportentry{FileUpload}{38}, \exportentry{SearchFilter}{38}. \\

\addlinespace[2pt]

\textsc{Naive-Ward} & T1 & 5 &
\exportentry{Header}{100}, \exportentry{loadStorage}{100}, \exportentry{saveStorage}{100}, \exportentry{Footer}{75}, \exportentry{FormField}{75}. \\

\textsc{Naive-Ward} & T2 & 8 &
\exportentry{Button}{62}, \exportentry{Footer}{62}, \exportentry{FormField}{62}, \exportentry{createAuthContext}{62}, \exportentry{Modal}{25}, \exportentry{useLocalStorage}{12}, \exportentry{getStorage}{0}, \exportentry{setStorage}{0}. \\

\textsc{Naive-Ward} & T3 & 6 &
\exportentry{loadJSON}{100}, \exportentry{saveJSON}{100}, \exportentry{Footer}{62}, \exportentry{ImageUpload}{38}, \exportentry{LoginForm}{38}, \exportentry{RegisterForm}{25}. \\

\addlinespace[2pt]

\textsc{SLA-Full} & T1 & 13 &
\exportentry{loadStorage}{100}, \exportentry{saveStorage}{100}, \exportentry{validateRequired}{100}, \exportentry{Header}{88}, \exportentry{MessageBanner}{88}, \exportentry{useFilteredList}{75}, \exportentry{Footer}{62}, \exportentry{Pill}{62}, \exportentry{EmptyState}{50}, \exportentry{TabBar}{38}, \exportentry{fileToDataURL}{38}, \exportentry{validateMinLength}{38}, \exportentry{renderPills}{12}. \\

\textsc{SLA-Full} & T2 & 17 &
\exportentry{useLocalStorage}{100}, \exportentry{useRouter}{100}, \exportentry{validateRequired}{100}, \exportentry{useFormError}{88}, \exportentry{FormField}{75}, \exportentry{NotificationBanner}{75}, \exportentry{Pill}{75}, \exportentry{validateAll}{75}, \exportentry{EmptyState}{62}, \exportentry{Footer}{50}, \exportentry{TabBar}{50}, \exportentry{useSearchFilter}{50}, \exportentry{usePhotoUpload}{38}, \exportentry{validateEmail}{38}, \exportentry{getTodayDate}{25}, \exportentry{useFieldErrors}{25}, \exportentry{filterBySearch}{12}. \\

\textsc{SLA-Full} & T3 & 11 &
\exportentry{FormField}{100}, \exportentry{Card}{88}, \exportentry{ErrorMessage}{88}, \exportentry{loadFromStorage}{88}, \exportentry{saveToStorage}{88}, \exportentry{usePersistentList}{88}, \exportentry{EmptyState}{75}, \exportentry{useFilteredList}{75}, \exportentry{Pill}{62}, \exportentry{Grid}{62}, \exportentry{useImageUpload}{38}. \\

\bottomrule
\end{tabular}
\caption{
Complete Super Library export inventory for WebGen-Bench Suite 3. 
Reuse rate denotes the percentage of the eight applications in the suite that import each symbol.
}
\label{tab:export-listing-suite3}
\end{table*}

\begin{figure*}[t]
  \centering
  \includegraphics[width=\textwidth]
  {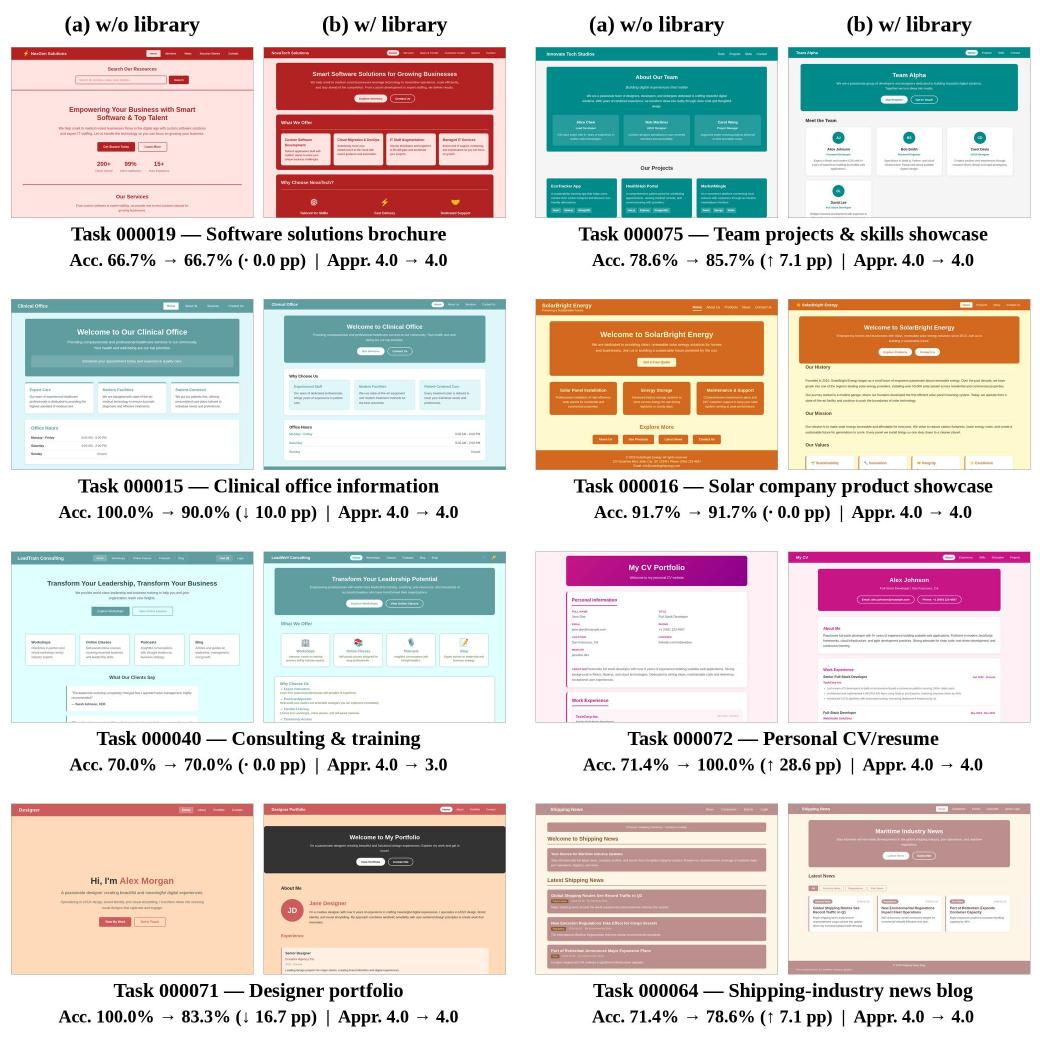}
  \caption{
\textbf{All eight Content-Presentation tasks under trial 1.}
Each task is shown as a (without library, with library) pair; the eight tasks are tiled as four rows of two tasks. UI-test accuracy and appearance grade are reported below each pair.
  }
  \label{fig:lib_prior_all_c2}
\end{figure*}


\section{Responsible NLP Checklist Details}
\label{appendix:responsible-nlp}

This appendix expands on the items listed in the Responsible NLP
Checklist that did not fit in the main text.

\subsection{Scientific Artifacts}
\label{appendix:responsible-nlp:artifacts}

\paragraph{Artifacts we use.}
Our experiments rely on the following external artifacts: the
\textbf{WebGen-Bench} benchmark~\citep{lu2025webgenbench}, the
\textbf{PaperBench} Code-Dev benchmark
~\citep{starace2025paperbenchevaluatingaisability}, the
\textbf{WebVoyager} UI agent used as WebGen-Bench's executable test
runner~\citep{he-etal-2024-webvoyager}, the
\textbf{mini-SWE-agent} harness~\citep{yang2024sweagent}, the
\textbf{cocoindex-code} code-indexing tool used by our Global Extract /
Apply candidate selectors, and the \textbf{Qwen-2.5-7B} reference model
used to compute MDL. We additionally call hosted LLM APIs
(\texttt{deepseek-v4-flash}, \texttt{gpt-5-mini},
\texttt{gpt-5.4-nano}, \texttt{Claude Opus 4.7}) for code generation
and judging. License information is summarized in
Table~\ref{tab:appendix:artifact-licenses}.

\begin{table}[h]
\centering
\footnotesize
\setlength{\tabcolsep}{4pt}
\begin{tabular}{lll}
\toprule
Artifact & License & Use \\
\midrule
WebGen-Bench       & MIT        & eval suite \\
PaperBench         & MIT        & eval suite \\
WebVoyager         & Apache 2.0 & UI runner  \\
mini-SWE-agent     & MIT        & harness    \\
cocoindex-code     & Apache 2.0 & indexing   \\
Qwen2.5-7B         & Apache 2.0 & MDL ref.   \\
\bottomrule
\end{tabular}
\caption{External artifacts and their licenses. All are permissive
open-source licenses, and our use is consistent with each artifact's
intended purpose.}
\label{tab:appendix:artifact-licenses}
\end{table}

\paragraph{Consistency with intended use.}
All listed artifacts are released for open use in research on automated
code generation, web agent evaluation, or language modeling. Our use
--- evaluating an LLM-based multi-agent coding system on the public
task splits, without redistributing benchmark data, grader rubrics, or
ground-truth code --- is consistent with each artifact's stated
intended use. We do not modify any benchmark task descriptions or
graders; we only adapt them into our \emph{suite} format by sampling
disjoint task subsequences (\S\ref{sec:experiments:benchmarks}).

\paragraph{Artifacts we create.}
The artifacts produced by this work --- the source code of the system,
the metric and evaluation result files behind every reported number, and
the final-round application codebases together with the Super Libraries
extracted from them --- are released under the MIT license. The generated codebases are
produced from benchmark task specifications and are intended as evidence
for the claims in this paper rather than as deployable applications.

\subsection{AI Assistants}
\label{appendix:responsible-nlp:ai-assistants}

We disclose the use of AI assistants in two distinct roles.

\paragraph{As research subjects.}
LLMs and LLM-based coding agents are the \emph{object of study} in
this paper. Specifically, \texttt{deepseek-v4-flash} is used as the
backbone for all compared agents, \texttt{gpt-5.4-nano} produces
per-block code summaries inside our pipeline,
\texttt{gpt-5-mini} serves as the appearance judge and the
report parser, and \texttt{Claude Opus 4.7} is used to render the
reference layout descriptions provided to all methods. These
usages are described in
\S\ref{sec:experiments:compared-methods} and
\S\ref{sec:experiments:metrics} and are integral to the experimental
setup, not authoring aids.

\paragraph{As research and engineering aids.}
During this project we used \textbf{ChatGPT (GPT-5 family)} and
\textbf{Claude (Opus 4.7)} as auxiliary tools in three bounded roles.
They sanity-checked the logical flow and consistency of arguments in
draft sections, stress-tested research ideas and surfaced
counter-examples or related work we might have missed, and assisted
with implementation-level code generation such as boilerplate analysis
scripts and plotting code.
All scientific claims, experimental designs, analyses, and final
interpretations are authored and verified by the human authors; AI
assistants did not autonomously produce any of the experimental
results, ablations, or conclusions reported in this paper.

\end{document}